\documentclass[final,1p,times,number]{elsarticle}
\usepackage{amssymb,amsmath}
\usepackage{mathbbol}
\usepackage{hyperref}
\journal{Computer Physics Communication}
\begin{document}
\begin{frontmatter}
\title{Effective noise reduction techniques for disconnected loops in Lattice QCD}
\author{Gunnar S.\ Bali}
\ead{gunnar.bali@physik.uni-regensburg.de}
\author{Sara Collins}
\ead{sara.collins@physik.uni-regensburg.de}
\author{Andreas Sch\"afer}
\ead{andreas.schaefer@physik.uni-regensburg.de}
\address{Institut f\"ur Theoretische Physik, Universit\"at Regensburg,
93040 Regensburg, Germany}
\begin{abstract}Many Lattice QCD observables of phenomenological
interest include so-called all-to-all propagators.
The computation of these requires prohibitively large computational
resources, unless they are estimated stochastically.
This is usually done. However, the computational demand can
often be further reduced by one order of magnitude by
implementing sophisticated unbiased noise reduction techniques. 
We combine both well known and
novel methods that can be applied to a wide range
of problems. We concentrate on calculating disconnected
contributions to nucleon structure functions, as one realistic
benchmark example.
In particular we determine the strangeness contributions
to the nucleon, $\langle N|\bar{s}s|N\rangle$ and to the spin of
the nucleon, $\Delta s$.
\end{abstract}

\end{frontmatter}
\section{Introduction}
Advanced Lattice QCD calculations often
require the evaluation of diagrams with
disconnected quark lines.
Important examples are properties of flavour singlet
mesons~\cite{Jansen:2008wv}, QCD spectroscopy including
multiquark and scattering states~\cite{Bulava:2008qx},
the determination of hadronic scattering lengths~\cite{Aoki:2007rd}, and
of isosinglet contributions to hadronic form factors and
structure functions~\cite{Zanotti:2008zm,Deka:2008xr}.
Moreover, statistical errors of some standard observables like
meson masses and electroweak decay constants can be reduced
by averaging the sources over the lattice volume.
In all these cases the standard point-to-all propagators that are
obtained by calculating twelve (colour times spin) columns of
the inverse lattice Dirac operator are not sufficient. Instead,
all-to-all or timeslice-to-all propagators need to
be computed.

Lattices typically contain a number of sites ranging from
$V\approx 10^6$ up to $V\approx 10^9$. Inverting
a lattice Dirac operator by conventional means would increase
the effort in terms of computer memory and operations spent
by this factor, relative to the cost of obtaining
a single point-to-all propagator. An alternative approach to the
problem consists of calculating an unbiased stochastic estimate of 
the propagator, replacing a factor $12V$ in effort by a
number $N$ of random sources, where in
certain situations $N$ can be as small as $10$. Such estimation
is permissible since, provided it is unbiased, it will only
add to the statistical errors. These are present in any case,
due to the path integral evaluation of expectation values of
observables by means of a Markov Monte Carlo simulation
(for an introduction into Lattice QCD simulations see e.g.\
\cite{DeGrand:2006zz}).

In general, the additional error introduced by this procedure
will, for sufficiently large numbers of estimates $N$,
decrease in proportion
to $1/\sqrt{N}$. Ideally, $N$ should be chosen such
that this additional error is
smaller than the statistical error induced by the Monte Carlo
time series. The optimal number will depend on the observable in
question and on the prescription employed to calculate
the stochastic estimate. In this article we discuss different
improvement methods aimed at decreasing the prefactor of the
asymptotic $1/\sqrt{N}$ behaviour. In particular, we will introduce
a new such method which we call the truncated solver method (TSM).
We then test combinations of different improved stochastic
estimator methods in a realistic Lattice QCD simulation.
As our benchmark examples we choose to calculate
the strangeness contribution to the spin
of the nucleon $\Delta s$ as well as the scalar strangeness
content $\langle N|\bar{s}s|N\rangle$.

The spin of the nucleon can be factorized into
a quark spin contribution
$\Delta\Sigma$, a quark angular momentum contribution $L_q$
and a gluonic contribution (spin and angular momentum) $\Delta G$:
\begin{equation}
\frac12=\frac12 \Delta\Sigma+L_q+\Delta G\,.
\end{equation}
In the na\"{\i}ve nonrelativistic
$\mathrm{SU}(6)$ quark model, $\Delta \Sigma=1$, with
vanishing angular momentum and gluon contributions. In this
case sea quark contributions will be absent too and
therefore there will be no strangeness
contribution $\Delta s$ in the factorization,
\begin{equation}
\Delta\Sigma=\Delta d+\Delta u +\Delta s+\cdots\,,
\end{equation}
where in our notation $\Delta q$ contains both, the spin of
the quarks $q$ and of the antiquarks $\bar{q}$.
Experimentally, $\Delta s$ is usually
obtained by integrating the strangeness contribution to
the spin structure function $g_1$ over the momentum fraction $x$.
The integral over the range in which data exists ($x\gtrsim 0.004$)
usually agrees with zero. For instance a recent Hermes measurement
in the region
$x\geq 0.02$ yields~\cite{Airapetian:2008qf} $\Delta s=0.037(19)(27)$.
This means that non-zero results
rely on extrapolations into the experimentally unprobed region of 
very small $x$ and are
model dependent~\cite{Zhu:2002tn,deFlorian:2009vb}.
The standard Hermes analysis~\cite{Airapetian:2007mh}
yields $\Delta s=-0.085(13)(8)(9)$
at a renormalization
scale $\mu^2=5\,$GeV${}^2$ in the $\overline{MS}$ scheme.
Our results below suggest,
$\Delta s=-0.017(21)$, unless there are large effects
in the chiral extrapolation from a pseudoscalar mass
$m_{\rm PS}\approx 450$~MeV
to the physical point. 

The scalar strangeness density is not directly accessible in
experiment but plays a r\^ole in models of nuclear structure.
It is also of phenomenological interest since,
assuming that heavy flavours are strongly suppressed,
the dominant coupling of the Higgs particle to the nucleon will
be accompanied by this scalar matrix element.

The outline of this article is as follows:
in Sec.~\ref{sec:red} we introduce our notation and
explain the stochastic estimator and noise reduction
methods applied. In Sec.~\ref{sec:discon} we detail
our lattice setup and employ these techniques to calculate
disconnected quark loop contributions. Finally, in Sec.~\ref{sec:result}
we present our results on the axial and scalar nucleon
strangeness matrix elements, before we conclude.
The systematic tuning of the parameters used in the 
truncated solver method is described in detail in \ref{sec:app1}.
The TSM is quite generally applicable.
This is also demonstrated in the appendix
where the conjugate gradient (CG) solver is replaced by the popular
stabilized biconjugate gradient
(BiCGstab2) solver~\cite{Frommer:1994vn}.
Preliminary results
appeared in~\cite{Collins:2007mh} and~\cite{Bali:2008sx}.

\section{Noise reduction techniques}
\label{sec:red}
As outlined above we require propagators $M^{-1}$ connecting arbitrary
pairs of lattice points, where $M$ denotes a lattice Dirac
operator. In our specific case we employ the Wilson quark
action~\cite{Wilson:1975vw,DeGrand:2006zz}.
Since the propagators have
$12 V\times 12V$ components, where $V$ denotes the number
of lattice points, direct
evaluation would be prohibitively expensive, both in terms of
computer time and of memory. However, we encounter statistical
errors anyway from the importance sampling of path
integral expectation values. Hence it is sufficient to aim
at an (unbiased) estimate, which can be obtained by stochastic
methods~\cite{Bitar:1988bb}.
For this purpose
we introduce the following notation,
\begin{equation}
\overline{A}=\overline{A}^N:=\frac{1}{N}\sum_{j=1}^NA^j\,.
\end{equation}
$N$ denotes the number of ``stochastic
estimates''.
Let $|\eta^i\rangle$, $i=1,\ldots,N$ be random vectors with the
properties,
\begin{align}
\overline{|\eta\rangle}&= {\mathcal O}\left(1/\sqrt{N}\right)\,,\\
\overline{|\eta\rangle\langle\eta|}&= \mathbb{1}
+{\mathcal O}\left(1/\sqrt{N}\right)\,.\label{eq:cancel}
\end{align}
These requirements are for instance met by complex
$\mathbb{Z}_2$ noise where
the $12V$ components are numbers $e^{i\phi}$, with the uncorrelated
random phases $\phi\in\{\pm\pi/4,\pm 3\pi/4\}$.
In \cite{z2min,Dong:1991xb,Bernardson:1993yg} it
has been demonstrated that real and complex
${\mathbb Z}_2$ noise reduces the variance, relative to other
choices such as Gaussian or double hump noise. In our
experience, similarly small variances
can also be obtained with ${\mathbb Z}_N$, $\mathrm{U}(1)$ or even
with $\mathrm{SU}(3)$ noisy sources~\cite{Babich:2007jg},
indicating that the only
important property is an equal modulus of the components. 
In the present study we employ complex ${\mathbb Z}_2$ noise,
where the components of the
random vectors $|\eta^i\rangle$
run over the spacetime volume, spin and colour.

We define the Hermitian lattice Dirac operator $Q=\gamma_5M$.
If we solve the linear systems,
\begin{equation}
\label{eq:lin}
Q|s^i\rangle=|\eta^i\rangle\,,
\end{equation}
for $|s^i\rangle$, then
we can substitute, see Eq.~(\ref{eq:cancel}),
\begin{equation}
\label{eq:set}
Q^{-1}=\overline{|s\rangle\langle\eta|}+
Q^{-1}\left({\mathbb 1}-\overline{|\eta\rangle\langle\eta|}\right)
\approx \mathrm{E}(Q^{-1}):=\overline{|s\rangle\langle\eta|}\,.
\end{equation}
The difference between the approximation of Eq.~(\ref{eq:set}) above
and
the exact result is unbiased and reduces like $1/\sqrt{N}$.
The sparse linear system of Eq.~(\ref{eq:lin}) can for example be
solved by means of the
conjugate gradient (CG) or the stabilized
biconjugate gradient (BiCGstab2) algorithms.
Note that in the CG case we can actually use even/odd preconditioning
by employing the operator
$M^{\dagger}M=Q^2$. We then obtain
$|s^{i}\rangle$ by multiplying the result with $Q$.

The stochastic estimate approach
reduces the problem from ${\mathcal O}(12V\times 12 V)$
to ${\mathcal O}(N\times 12 V)$, in terms of memory and computer time.
The stochastic error will remain roughly constant if $N$ is
scaled like $\sqrt{V}a^2$ with the lattice volume, where
$a$ denotes the lattice spacing.
In order to limit the computational effort,
$N$ should not be chosen overly large.
However, in the end the stochastic noise should
not be the dominant source of statistical error.
In general,
the optimal balance between the stochastic sampling
on single configurations and the Monte Carlo importance sampling
of gauge configurations will depend
on the observable
in question and on the methods used.

Due to the difference between $\overline{|s\rangle\langle\eta|}$ and $Q^{-1}$
above, any fermionic observable $A$ can only be estimated up to a
stochastic error $\Delta_{A, \rm stoch}=\mathcal{O}(1/\sqrt{N})$ on a
given configuration. We define the configuration average
$\langle\cdot\rangle_c$ over $n$ uncorrelated
configurations and normalize this appropriately:
\begin{equation}
\label{eq:sto}
\sigma_{A,\rm stoch}^2:=
\frac{\langle\Delta_{A,\rm stoch}^2\rangle_c}{n}\,.
\end{equation}
For large $N$ and $n$ this will scale like
$\sigma_{A,\rm stoch}^2\propto (Nn)^{-1}$.
We also define the total error
$\sigma_{A,\rm tot}$
as the variation of the obtained estimates of $A$ over gauge
configurations. At fixed $N$ this will be proportional
to $1/\sqrt{n}$.
Obviously, this total error is limited by
the stochastic error\footnote{For an exact calculation of $A$
($\sigma_{A,\rm stoch}=0$), a Monte Carlo error
$\sigma_{A,\rm stat}=\sigma_{A,\rm tot}\propto 1/\sqrt{n}$ can be
introduced. In the limit of large $N$ and $n$, the factorization
$\sigma_{A,\rm stat}^2=\sigma_{A,\rm tot}^2-\sigma_{A,\rm stoch}^2$ holds,
see e.g.~\cite{Wilcox:1999ab}. For our considerations there is no
need to introduce this quantity.}:
\begin{equation}
\label{eq:MCE}
\sigma^2_{A,\rm tot}>\sigma^2_{A,\rm stoch}\,.
\end{equation}
If $\sigma^2_{A,\rm stoch}\simeq\sigma^2_{A,\rm tot}$
then it is worthwhile to improve the quality
of the estimates while if $\sigma^2_{A,\rm stoch}\ll
\sigma^2_{A,\rm tot}$ then precision can only
be gained by increasing the number of configurations
$n$, possibly
reducing $N$ to save computer time since the
$n^{-1}$ scaling cancels from the above inequality.

To minimize the stochastic noise at a given computational effort,
we combine a multitude of techniques:
\begin{enumerate}
\item partitioning (also known as the spin explicit method or as
dilution)~\cite{Bernardson:1993yg,Viehoff:1997wi,Foley:2005ac},
\item the hopping parameter expansion (HPE) technique~\cite{Thron:1997iy,Michael:1999rs,Bali:2005pb},
\item the truncated solver method (TSM)~\cite{Collins:2007mh} and
\item the truncated eigenmode acceleration (TEA)~\cite{DeGrand:2002gm,Bali:2005pb,Foley:2005ac}.
\end{enumerate}
These methods are explained below.

\subsection{Partitioning}
\label{sec:part}
We decompose
${\mathcal R}=V\,\otimes\,\mbox{colour}\,\otimes\,\mbox{spin}$
into $m$ subspaces ${\mathcal R}_j$:
${\mathcal R}=\oplus_{j=1}^m{\mathcal R_j}$.
One can then set the source vectors $|\eta^i_{|j}\rangle$
to zero, outside of the subspace ${\mathcal R_j}$. We label
the corresponding solutions as $|s^i_{|j}\rangle$.
The solution for the all-to-all propagator is then given by
the sum,
\begin{equation}
Q^{-1}\approx \sum_{j=1}^m \overline{|s_{|j}\rangle\langle
\eta_{|j}|}\,.
\end{equation}
This procedure results in an $m$-fold increase in the total number of
solver applications. The term
${\mathbb 1}-\overline{|\eta\rangle\langle\eta|}$ multiplying $Q^{-1}$
in Eq.~(\ref{eq:set}) only has off-diagonal entries, all
of similar sizes. Therefore the off-diagonal entries
of $Q^{-1}$ will determine the stochastic error of
the particular observables in question.
$Q^{-1}$ will exponentially decay with the spacetime distance
between source and sink: 
spacetime components in the neighbourhood
of the sink position will yield 
the leading contributions to the stochastic variance.
Likewise the noise for spin components that are strongly coupled
to each other by $Q^{-1}$, multiplied by the relevant $\Gamma$
and derivative structures, will dominate.
Ideally, partitioning will black out the largest
contributions to the stochastic noise.
If the achieved reduction exceeds the $1/\sqrt{m}$ factor, then the
computational overhead is justified.
This overhead can be somewhat reduced at
the expense
of memory by sophisticated preconditioning techniques
and/or
aggressive deflation~\cite{Morgan:2004zh,Luscher:2007se,Stathopoulos:2007zi,Brannick:2007ue,Wilcox:2007ei,Tadano:2009gg}: the constant
set-up costs of
such techniques
become more affordable with a larger number of right hand sides.
We experimented with various patterns and
combinations of
colour, spin and time partitioning, see e.g.~\cite{Ehmann:2009ki}.
Partitioning
as a stand-alone solution works very well
in many situations~\cite{Viehoff:1997wi,Gusken:1999te,Wilcox:1999ab,Foley:2005ac}.
However, such gains are mostly eliminated once partitioning is
combined with the other tricks.
Time partitioning is a
notable exception: in situations where only 
timeslice-to-all
propagators are required there is no increase in the
number of solves but the variance is still reduced.
In the calculation of disconnected contributions to the
nucleon structure it also turns out to be useful to
generate the current insertion at more than one timeslice,
e.g.\ to average
the correlation with a nucleon
propagating from a given source in the forward direction and
the backward propagating one (which comes for free).
In this case one can seed the random sources at
two (or more) well separated timeslices and still gain
from partitioning, without any associated overhead.

\subsection{The hopping parameter expansion technique}
The stochastic noise from terms that are
close to the diagonal is accompanied
by larger amplitudes than terms that are far off the
diagonal, see Eq.~(\ref{eq:set})
and the discussion in Sec.~\ref{sec:part} above.
Hence the cancellation of near-diagonal noise
requires a comparatively larger number of estimates.
The HPE aims at
eliminating some of the near-diagonal noise contributions
by exploiting the ultra-locality of the action.
Thus it cannot be generalized for instance to
the Neuberger action~\cite{Neuberger:1997fp}. 

We rewrite the fermionic matrix as,
\begin{equation}
\label{eq:kappad}
2\kappa M={\mathbb 1}-\kappa D\,.
\end{equation}
The HPE is based on the observation that
one can expand,
\begin{equation}
M^{-1}=2\kappa\sum_{i=0}^{\infty}(\kappa D)^i=
2\kappa\sum_{i=0}^{k-1}(\kappa D)^i+(\kappa D)^kM^{-1}\,,
\label{eq:hpe2}
\end{equation}
where $k\geq 1$. For distances between source and
sink consisting of more than $k$ links,
the first term on the right hand side does not
contribute since $D$ only connects nearest
spacetime neighbours. Therefore, $M^{-1}_{xy}=
[(\kappa D)^kM^{-1}]_{xy}$ for sufficiently large source and
sink separations. However, at finite $N$, $\{\mathrm{E}[M^{-1}]\}_{xy}\neq
\{\mathrm{E}[(\kappa D)^kM^{-1}]\}_{xy}
(=\{(\kappa D)^k\mathrm{E}[M^{-1}])\}_{xy})$,
where the variance of the latter
estimate of $M^{-1}_{xy}$ is reduced. This was for instance exploited
in~\cite{Bali:2005pb}.
In some cases additional powers of $D$ can be gained due to the
$\Gamma$ structure of the creation and annihilation operators.

Here we study closed loops, i.e.\ $x=y$.
Obviously, only even
powers of $D$ contribute to
$\mathrm{Tr}\,(M^{-1}\Gamma)$, where $\Gamma\in\{{\mathbb 1},
\gamma_{\mu},\sigma_{\mu\nu},\gamma_{\mu}\gamma_5,\gamma_5\}$.
We can write,
$\mathrm{Tr}\,(M^{-1}{\mathbb 1})
=2\kappa\,\mathrm{Tr}\,{\mathbb 1}+
\kappa^k\mathrm{Tr}\,(D^kM^{-1})$,
where the first term
can be computed trivially. In the other cases,
$\mathrm{Tr}\,(M^{-1}\Gamma)=\kappa^k\mathrm{Tr}\,(D^kM^{-1}\Gamma)$.
For the Wilson action, $k=8$ for $\Gamma\in\{\gamma_{\mu}\gamma_5,\gamma_5\}$
and $k=4$ otherwise.
For the Sheikholeslami-Wohlert action~\cite{Sheikholeslami:1985ij},
$k=2$. The lowest non-vanishing terms have been calculated
analytically~\cite{Thron:1997iy} and can be
computed and corrected for exactly (unbiased noise subtraction)~\cite{Thron:1997iy,Mathur:2002sf,Deka:2008xr}.
We limit ourselves to the highest vanishing order 
($k=4$ or $k=8$ for the Wilson action), where such subtraction is
not necessary. In this case
the computational overhead of the HPE is
small, such that this substitution can easily be combined with the
TSM that we explain below.

\subsection{The truncated solver method}
\label{sec:tsm}
It has been noticed long ago~\cite{Eicker:1996gk} that solvers
typically converge to the correct result within an
accuracy of the size of the stochastic error after a
relatively small number of iterations. Practitioners
have therefore sometimes relaxed the requirement on the residual
when solving for stochastic sources. This is not
unproblematic
since it introduces a systematic bias that will
be invisible on one configuration but might very well
affect the result obtained on a sample of, say, 200 configurations.
We label the result obtained after $n_{\rm t}$ solver iterations by
$|s_{(n_{\rm t})}^i\rangle$, where omitting the subscript means
convergence within numerical accuracy.
We can now factorize:
\begin{equation}
Q^{-1}\approx  \mathrm{E}(Q^{-1}):=\frac{1}{N_1}\sum_{i=1}^{N_1}
|s^i_{(n_{\rm t})}\rangle\langle \eta^i|
+\frac{1}{N_2}\sum_{i=N_1+1}^{N_1+N_2}
\left(|s^i\rangle-|s^i_{(n_{\rm t})}\rangle\right)\langle\eta^i|\,.
\label{trunc}
\end{equation}
The above equation is an exact linear
decomposition of $Q^{-1}$ and the algorithm
used to calculate both parts is well defined.
Due to these properties and the fact that the $|\eta^i\rangle$
with $i\leq N_1$ are uncorrelated with those
for $i>N_1$, the resulting estimate is unbiased.
Ideally, one will generate a large number $N_1\gg N_2$
of relatively cheap estimates at small $n_{\rm t}$ and
then remove the bias by correcting with a small
number $N_2$ of expensive solutions to machine precision.
This method can easily be combined
with all the other methods described here. It is possible
to tune the two parameters $N_1/N_2$ and $n_{\rm t}$ that
enter the algorithm
to the point of minimal variance at fixed computer time in a relatively
inexpensive and straightforward way.
This is discussed in \ref{sec:app1}.

In some sense the underlying philosophy of TSM is similar to
estimating the cheap first term within the HPE
Eq.~(\ref{eq:hpe2}) with many random sources and the
expensive second term with only a few sources. However,
iterative solvers like the CG converge faster than the HPE.
Moreover, TSM is applicable to any fermion action, not only to
ultra-local ones, and efficient for any quark mass. 
TSM can be combined with HPE,
see Sec.~\ref{sec:combine} below.

To check our implementation of the method we compared the exact result for
$(M^{-1})_{x,y}^{s_1c_1,s_2c_2}$, where $s_1c_1$ denotes the spin
and colour indices, $x=(0,0,0,3)$ and $y=(i,0,0,3)$, $i=0\ldots 10$,
with an
estimate obtained from Eq.~(\ref{trunc}). We  indeed find
consistency within errors for different $n_{\rm t}$, $N_1$ and
$N_2$.  For example, for $n_{\rm t}=5$,
$N_1=5500$, $N_2=300$, $i=1$, $s_1=s_2=1$,
$c_1=c_2=2$, $\mathrm{E}[M^{-1}_{n_c}]=(0.0300(7),-0.0014(7))$,
compared to the exact result of $(0.0302\ldots,-0.0010\ldots)$.

In \ref{sec:app1} we also demonstrate that the TSM can be
generalized to other solvers, by employing BiCGstab2.
However, the smooth convergence of CG, that is relatively
independent of the random source and gauge configuration,
is a clear advantage when it comes to deciding on a TSM
parameter set.
Moreover, in the CG case the combination with the truncated
eigenmode acceleration bears less computational overhead
and is compatible with even-odd preconditioning (see below).
We emphasize however that the converged solutions
$|s^i\rangle$ within Eq.~(\ref{trunc}) can be obtained
by using any efficient solver. In particular, there is no need to employ the
same solver as that used for obtaining the
truncated solutions $|s^i_{(n_t)}\rangle$. The truncated solver
of course needs to be the same for $i>N_1$ as that used for $i\leq N_1$.

\subsection{The truncated eigenmode acceleration}
We define the
$n_{\rm e}$ smallest (real) eigenvalues $q_i$ of $Q$ and the corresponding
orthonormal eigenvectors $|u_i\rangle$, $i=1,\ldots, n$:
\begin{equation}
Q|u_i\rangle=q_i|u_i\rangle\quad,\quad
\langle u_i|u_j\rangle=\delta_{ij}\,.
\end{equation}
These can for instance
be calculated by means of the parallel implicitly restarted
Arnoldi method (IRAM) with Chebychev acceleration~\cite{Neff:2001zr}
or by Ritz methods~\cite{Kalkreuter:1995mm}.

We can now approximate~\cite{Neff:2001zr},
\begin{equation}
\label{eq:tea1}
Q^{-1}\approx \sum_{i=1}^{n_{\rm e}}|u_i\rangle q_i^{-1}\langle u_i|\,.
\end{equation}
However, this estimate is biased.
We define the projection
operator ${\mathbb P}_{n_{\rm e}}$, onto the complement of the space
spanned by these $n_{\rm e}$ eigenvectors,
\begin{equation}
\label{eq:project}
{\mathbb P}_{n_{\rm e}}={\mathbb 1}-\sum_{i=1}^{n_{\rm e}}|u_i\rangle\langle u_i|\,.
\end{equation}
Projecting the stochastic sources onto the orthogonal complement,
\begin{equation}
\label{eq:pro}
Q|s^i_{\perp}\rangle=|\eta^i_{\perp}\rangle:=
{\mathbb P}_{n_{\rm e}}|\eta^i\rangle\,,
\end{equation}
yields the new unbiased estimate~\cite{Bali:2005pb},
\begin{equation}
Q^{-1}\approx\mathrm{E}(Q^{-1}):=\sum_{i=1}^{n_{\rm e}}|u_i\rangle q_i^{-1}\langle u_i|
+\overline{|s_{\perp}\rangle\langle
\eta_{\perp}|}\,,
\label{eq:tea}
\end{equation}
where a left projection of $|s_{\perp}^i\rangle$ is not necessary
since $[Q,{\mathbb P}_{n_{\rm e}}]=0$ and thus
${\mathbb P}_{n_{\rm e}}Q\,{\mathbb P}_{n_{\rm e}}=Q\,{\mathbb P}_{n_{\rm e}}$.
The above estimate will usually have a reduced variance.
Note that in the literature exact point-to-all propagators have
also been combined with a truncated low eigenmode
all-to-all calculation
to achieve smaller gauge errors (low mode
averaging)~\cite{DeGrand:2004qw,Giusti:2004yp}.

A nice side effect of TEA lies in the acceleration of the solver,
by deflation~\cite{deForcrand:1995bs,Morgan:2004zh,Stathopoulos:2007zi,Wilcox:2007ei,Darnell:2007dr,Tadano:2009gg}.
The condition number of the projected operator
$Q\,{\mathbb P}_{n_{\rm e}}$
and therefore the number of solver iterations are reduced.
Within some algorithms like the CG
the fact that the projector commutes with the operator
$Q$ guarantees that
the Krylov subspace remains confined within the
orthogonal complement. So in these cases
no further intermediate projections are necessary to fully
exploit the potential of deflation.

Note that while
the number of undeflated solver iterations increases like
$1/m_{\rm PS}^2$ at small quark masses,
the efficiency of the eigenvalue calculation remains the same.
Unfortunately, at small $m_{\rm PS}$, one would like to
increase the linear lattice extent $L\propto 1/m_{\rm PS}$.
In this case,
the rank of the deflation space in the worst case
will increase like
$n_{\rm e}\propto L^4\propto Va^4$.
Depending on the volume required this may or may not become
a serious problem.
The volume scaling of stochastic methods is somewhat more favourable:
the number of estimates needs to be adjusted at most like
$\sqrt{V}a^2$ (this can be somewhat reduced by partitioning).
{}From these considerations it becomes evident that the optimal
choices of $n_{\rm e}$ and $N$ strongly depend on the
volume, the quark mass and
the lattice spacing. The same holds for the algorithm where a deflated CG
can outperform a (deflated) BiCGstab2, in particular
when combined with the truncated solver method.

Obviously, it is also possible to decompose $M$, rather than
the Hermitian operator $Q=\gamma_5M$ of Eq.~(\ref{eq:tea1}), into
eigenmodes~\cite{Hip:2001hc}:
$M|r_i\rangle=\lambda_i|r_i\rangle$. However, in this case
we end up with a biorthonormal system $\langle\ell_i|r_j\rangle=
\delta_{ij}$, where
the left eigenvectors $|\ell_i\rangle$ will differ from the
right eigenvectors:
$\langle \ell_i|M=\langle\ell_i|\lambda_i$.
One can now decompose,
\begin{equation}
M^{-1}\approx \sum_{i=1}^{n_{\rm e}}|r_i\rangle \lambda_i^{-1}\langle \ell_i|\,.
\end{equation}
It can easily be seen that due to the property
$M^{\dagger}=\gamma_5M\gamma_5$,
$\tilde{\lambda}_i:=\lambda_i^*$ is an eigenvalue whenever
$\lambda_i$ is an eigenvalue,
with a left eigenvector $\langle \tilde{\ell}_i|=\langle r_i|\gamma_5$ and
a right eigenvector $|\tilde{r}_i\rangle=\gamma_5|\ell_i\rangle$.
The advantage of this decomposition is that 
the eigenvectors
are independent of the quark mass and the eigenvalues
at different $\kappa$ values are trivially related to each other.
This follows from the
structure Eq.~(\ref{eq:kappad}).
In this article we will not pursue this alternative
eigenmode decomposition any further.
However, the decomposition of $M^{-1}$ might converge better in some channels
than that of $Q^{-1}$ and vice versa.
While this biorthogonal approach is incompatible with deflating the CG solver,
it would be the natural starting point for BiCG solvers.

\subsection{Combining the methods}
\label{sec:combine}
Partitioning can trivially be combined with the other three
methods. However some notes are in place with respect to
combinations of HPE, TSM and TEA.
Within the HPE a left-multiplication of the estimate
$\mathrm{E}[Q^{-1}]$ with $D^k$ is essential
since we have defined $M^{-1}=Q^{-1}\gamma_5$.
A right-multiplication would involve $\gamma_5D\gamma_5=D^{\dagger}$.
The easiest way of implementing HPE is to multiply
the solution vectors with $(\kappa D)^k$, prior to any
application of smearing functions or contractions but after
TEA and TSM.

Within Eq.~(\ref{eq:tea}) only the projected source vectors
appear. Hence after the projection,
Eq.~(\ref{eq:pro}),
\begin{equation}
|\eta^i_{\perp}\rangle=|\eta^i\rangle-\sum_{j=1}^{n_{\rm e}}
\langle u_j|\eta^i\rangle\,|u_j\rangle\,,
\end{equation}
the original noise vectors $|\eta^i\rangle$ can be discarded.

The overhead from the projection can be significant
when combined with the TSM, Eq.~(\ref{trunc}), where the
large number of estimates
$N_1$ implies a large number of projections and the small 
number of solver iterations $n_{\rm t}$ indicates that not much
computer time will be spent between these projections.
The projection overhead can be reduced by restricting 
the computation of the inner products
to the partitioning subspace:
many components of
$|\eta^i\rangle$ might have been set to zero if the
partitioning method was used. This saving cannot be obtained at the sink.
Fortunately, $[{\mathbb P}_{n_{\rm ev}},Q]=0$, such that the solutions
$|s^i_{\perp}\rangle$
remain within the orthogonal complement and no
second projection is necessary, provided the residual
of the solver is chosen sufficiently small!

\begin{figure}
\centerline{\includegraphics[width=.9\textwidth,clip]{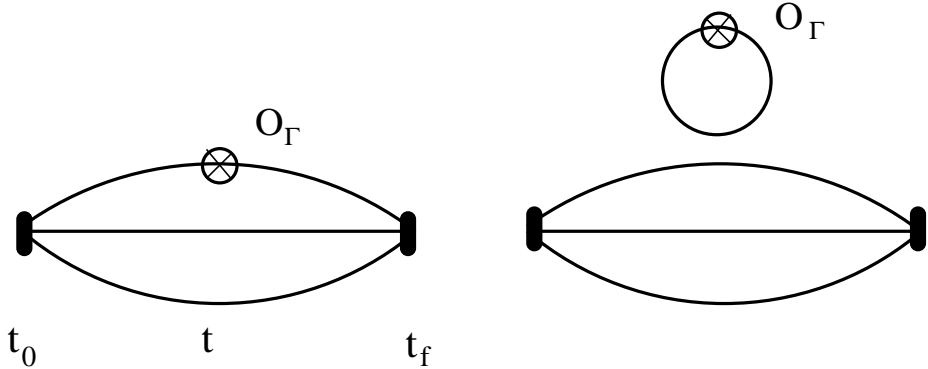}}
\caption{Connected and quark-line disconnected current insertion
into the nucleon.\label{discon_con}}
\end{figure}
Obviously the residual will not be small
for the truncated solutions, which therefore
one would wish to project back into the orthogonal complement
(before application of the $D$ operator within the HPE). 
Fortunately, the (even-odd preconditioned) CG algorithm
never leaves the orthogonal subspace: in this case, such an
additional projection is never necessary, even if the solver
is not run to convergence. Neither is such a left projection
strictly required for solvers without
this property, as long
as the precision of the part that is run to convergence is
sufficiently large. In this case, the $N_1$ biased estimates
will pick up some unwanted low mode contributions that will
however be corrected for by the $N_2$ estimates of the bias.
This might or might not increase the stochastic errors.

For completeness we mention that 
domain decomposition techniques have been suggested in the
literature~\cite{Michael:1998sg,Burch:2006mb}. These can
easily be combined with TSM and TEA too, as
can the so-called
one-end-trick~\cite{Foster:1998vw,McNeile:2006bz,Boyle:2008yd}.
We remark that if a bosonic representation is
employed~\cite{Michael:1998sg,Duncan:2001ta},
estimating $M^\dagger M=Q^2$
instead of $Q$, then the TSM can in principle
be substituted by
a quasi heatbath strategy~\cite{deForcrand:1998je}.

\section{Evaluation of the disconnected loop}
\label{sec:discon}
\subsection{Lattice setup}
We study combinations of the variance reduction
techniques outlined above,
using configurations provided by the Wuppertal group: these
are $n_{\rm f}=2+1$ dynamical configurations with $V=16^3\times 32$
lattice points, generated using a Symanzik
improved gauge action and a stout-link improved (rooted) staggered fermion
action. The lattice spacing is fairly coarse, $a^{-1}\approx 1.55$~GeV,
while the spatial extent is around $2$~fm. Further details can be found
in~\cite{latdetails}. For the valence quarks we use the Wilson action
with $\kappa=0.166$, $0.1675$ and $0.1684$, corresponding to
pseudoscalar masses of about $600$, $450$ and $300$~MeV, respectively.
The analysis is performed on 326 configurations at $\kappa_{\rm
  loop}=0.166$, 167 configurations at $\kappa_{\rm loop}=0.1675$ and
$152$ configurations at $\kappa_{\rm loop}=0.1684$, where $\kappa_{\rm
  loop}$ refers to the $\kappa$ value of the disconnected loop. Our
main results are obtained using the CG solver with
even-odd preconditioning.
However, results obtained with BiCGstab2
are given in \ref{sec:app1}. The code used
throughout is a modified version of
Chroma~\cite{chroma1,chroma2,chroma3}.

\subsection{Results for the disconnected loop}
We wish to calculate
nucleon structure observables. For this purpose a
nucleon will be created at some initial time $t_0$ and
destroyed at a later time $t_{\rm f}\gg t_0$, with a current
inserted at some intermediate time $t$. This is
illustrated in Fig.~\ref{discon_con}. The disconnected
loop $\mathrm{Tr}\,(M^{-1}\Gamma)$ within the right diagram
of the figure will be calculated with
stochastic all-to-all techniques and can subsequently be combined
with the nucleon two-point function, calculated in the
standard point-to-all way, on a configuration
by configuration basis. 

\begin{figure}
\centerline{\rotatebox{270}{\includegraphics[height=.471\textwidth,clip]{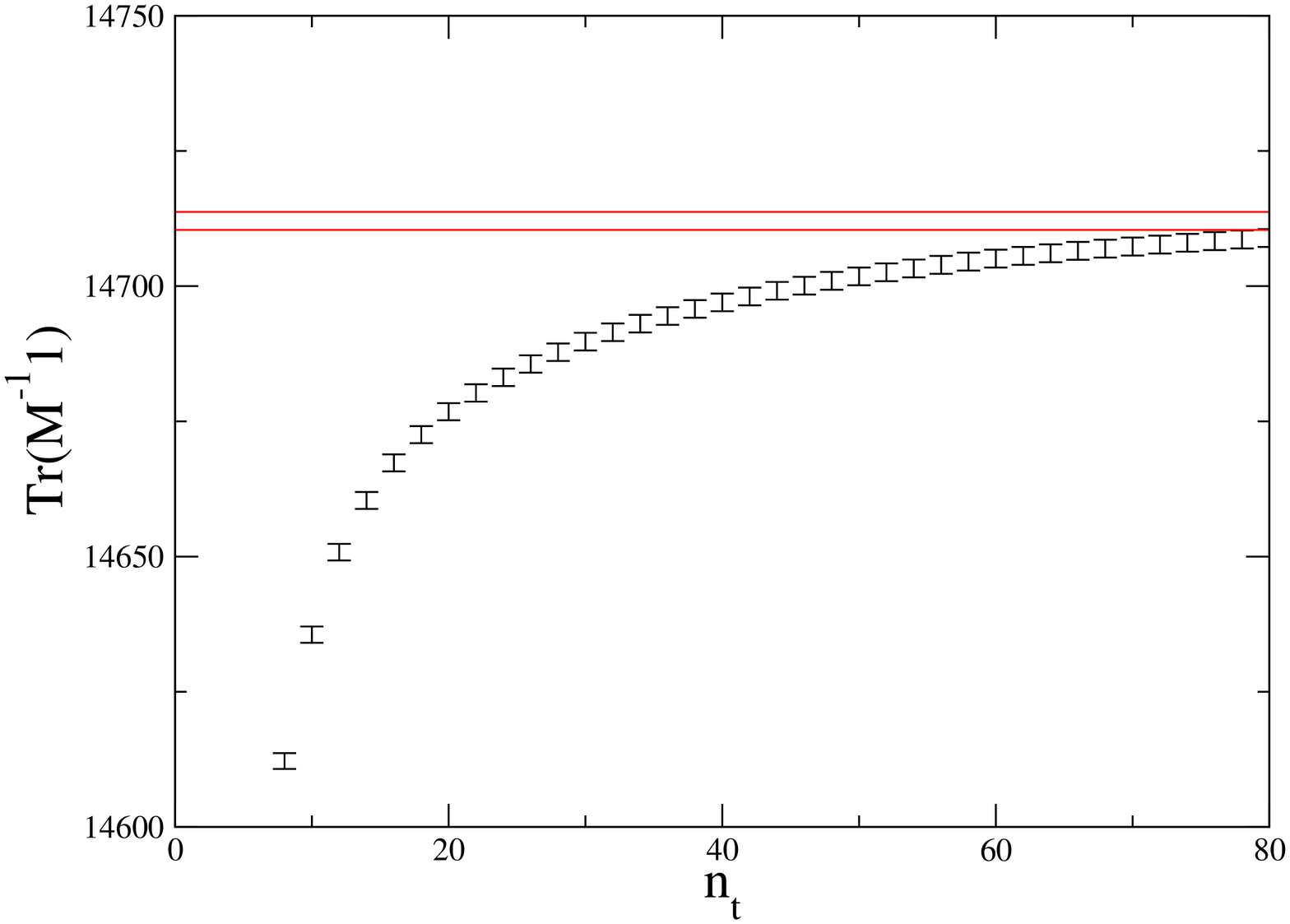}}\hspace{.02\textwidth}\rotatebox{270}{\includegraphics[height=.485\textwidth,clip]{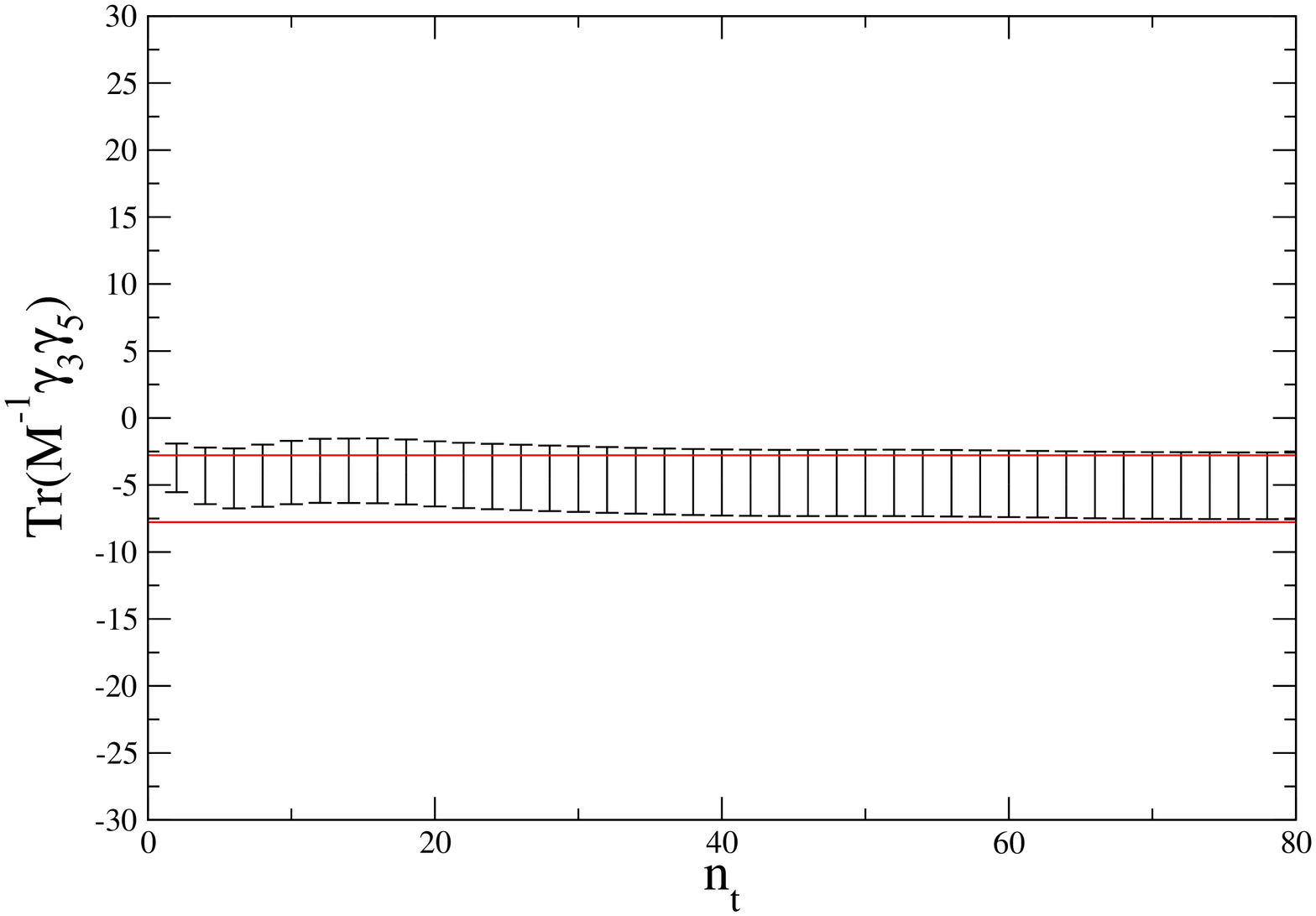}}}
\caption{Truncated estimates of the zero momentum projected
$\mathrm{Tr}\,(M^{-1}\Gamma)$, obtained
after $n_{\rm t}$ CG solver iterations for $\Gamma={\mathbb 1}$ (left)
and  for $\Gamma=\gamma_3\gamma_5$~(right) at $\kappa_{\rm loop}=0.166$.
The results are averaged over $300$ stochastic sources. Horizontal
lines indicate the result with statistical errors 
at convergence.\label{converge1}}
\end{figure}
A preliminary picture of how well the variance reduction techniques
work can be obtained by studying the zero-momentum
projected disconnected loop $\sum_{\mathbf x}\mathrm{Tr}\,
(M^{-1}_{xx}\Gamma)$ alone, where $x=(\mathbf{x},t)$.
With the exception of $\Gamma={\mathbb 1}$,
expectation values of these loops
over many configurations vanish, due to the discrete
charge and parity symmetries of QCD. However, the expectation
value of the correlation between loop and
proton (with a momentum injected or
with a specified helicity) can be non-zero. Likewise,
with the exception of
trivial cases such as $\mathrm{Im}\,\mathrm{Tr}\,(M^{-1}\gamma_5)=0$,
the loops will not vanish on single configurations. Using
the Euclidean spacetime convention
$\{\gamma_{\mu},\gamma_{\nu}\}=2\delta_{\mu\nu}$,
$\gamma_5=\gamma_1\gamma_2\gamma_3\gamma_4$,
we are interested in evaluating
$\mathrm{Re}\,\mathrm{Tr}\,(M^{-1}{\mathbb 1})$,
$\mathrm{Im}\,\mathrm{Tr}\,(M^{-1}\gamma_{\mu})$,
$\mathrm{Im}\,\mathrm{Tr}\,(M^{-1}\sigma_{\mu\nu})=
\mathrm{Re}\,\mathrm{Tr}\,(M^{-1}\gamma_{\mu}\gamma_{\nu})$
for $\mu\neq\nu$,
$\mathrm{Re}\,\mathrm{Tr}\,(M^{-1}\gamma_{\mu}\gamma_5)$
and
$\mathrm{Im}\,\mathrm{Tr}\,(M^{-1}\gamma_5)$.

We investigate the reduction in computer time of
optimized stochastic
estimates, relative to the situation without the use
of any improvement techniques~(except for time partitioning:
all disconnected loops are
calculated on one timeslice only).  We state all costs in terms
of the average {\em real} computer time required on a Pentium 4 PC for
one solver application (unimproved estimate), where we account for all
overheads of the improvement methods. We also ran the parallelized
code on IBM p6 and Opteron clusters as well as on BlueGene/L and BlueGene/P
systems.
The overall efficiencies of matrix-vector multiplies and global sums depend
on the architecture, however, the gain factors defined below
are not significantly affected.

We define the gain as the ratio,
\begin{equation}
\label{eq:gain}
\mathrm{gain}(\Gamma) =
\frac{\mathrm{var}
\{\mathrm{E}_0[\mathrm{Tr}\,(M^{-1}\Gamma)]\}}
{\mathrm{var}
\{\mathrm{E}_{\rm imp}[\mathrm{Tr}\,(M^{-1}\Gamma)]\}}\,,
\end{equation}
taken at fixed real computer time.
$\mathrm{var}$ denotes the variance between stochastic
estimates on a given gauge configuration,
$\mathrm{E}_0$ and
$\mathrm{E}_{\rm imp}$ stand for the unimproved and
improved estimated, respectively.

We start by investigating the TSM 
at the heaviest quark mass, $\kappa_{\rm loop}=0.166$.
Within TSM the combination
$\mathrm{Tr}\,(M^{-1}\Gamma)$ is obtained as an average
over $N_1$ truncated solutions $|s^i_{(n_{\rm t})}\rangle$:
 $\langle\eta^i|\gamma_5\Gamma|s^i_{(n_{\rm t})}\rangle$.
This value is then corrected by $N_2\ll N_1$
estimates of the resulting bias, see Eq.~(\ref{trunc}).
The faster 
the truncated estimate as a function of $n_{\rm t}$
approaches the estimate
obtained at full convergence, the better the method will work.
In this study we only consider
currents of local quark bilinears, where
we use the $\Gamma$ conventions of~\cite{chroma1}.
We observe very satisfactory convergence rates
for all the 16 possible $\Gamma$ structures.
In Fig.~\ref{converge1} we illustrate this for the
even/odd preconditioned CG solver for the
worst case ($\Gamma=\mathbb{1}$) and for the
best case ($\Gamma=\gamma_3\gamma_5$).
Note, however, that the unimproved estimate of
$\Gamma=\mathbb{1}$ is already very precise to start with.
Convergence at this
$\kappa$ value is reached after
$n_{\rm conv}\approx 480$ CG iterations.

\begin{table}
\begin{center}
\begin{tabular}{l|rrrrr|rrrrr}\hline
  $\mathrm{E}_{\rm imp}[\mathrm{Tr}\,(\mathrm{M}^{-1}\Gamma)]$
  &\multicolumn{5}{c|}{TSM} &\multicolumn{5}{c}{TSM+HPE}\\\hline
  $\Gamma$ \hfill & $\mathbb{1}$ \hfill & $\gamma_3$ \hfill & $\gamma_1\gamma_2$ \hfill &
  $\gamma_5$ \hfill & $\gamma_3\gamma_5$ \hfill& $\mathbb{1}$\hfill & $\gamma_3$ \hfill &
  $\gamma_1\gamma_2$ \hfill & $\gamma_5$ \hfill & $\gamma_3\gamma_5$\hfill\\\hline
  $n_{\rm t}$ & 50 & 27 & 14  & 18 & 18& 66 & 78 & 50  & 78 & 90\\
  $N_1/N_2$ & 23 & 21 & 32 & 28 & 30 & 26 & 25 & 21 & 26 & 26 \\
  $k$ & & & & & & 4 & 4 & 4 & 8 & 8 \\\hline gain & 5 & 5 & 10 & 8 &
  8& 8 & 11 & 19 & 25 & 30\\\hline
\end{tabular}
\end{center}
\caption{Optimized TSM values for $n_{\rm t}$ and
  $N_1/N_2$, see Eq.~(\protect\ref{trunc}),
at $\kappa=0.166$ for a subset of the $\Gamma$s studied,
calculated on one configuration using 300 estimates.
The approximate gains obtained at fixed cost, using these
values, are also shown. Where our method is
  combined with the HPE technique, $k$ indicates the order
  used.\label{table:optvalues1}}\end{table}

Applying TSM involves making choices for $n_{\rm t}$ and
for $N_1/N_2$. We detail our
optimization procedure in \ref{sec:app1}.
This systematic tuning results in similar amounts of
computer time being spent on the truncated estimate
as on estimating the bias. In particular we find
$N_1/N_2\approx
n_{\rm conv}/n_{\rm t}$.
Performing this optimization on a single configuration
appears sufficient since the variance is not much affected
if $n_{\rm t}$ and $N_1/N_2$
are changed by $20\,\%$. Fluctuations between
configurations turn out to be smaller than this value.

In Table~\ref{table:optvalues1} we display the
optimized parameters at $\kappa_{\rm loop}=0.166$
for a representative choice of
$\Gamma$ structures (scalar, vector,
tensor, pseudoscalar and axial vector),
together with the approximate
fixed cost gains, Eq.~(\ref{eq:gain}).
These factors vary between 5 and 10.
In a real production run one would not wish to
generate new sets of optimized estimates for each
$\Gamma$ structure or observable one is interested in.
Fortunately, with the still tolerable exception
of $\Gamma={\mathbb 1}$, $n_{\rm t}$ and $N_1/N_2$ exhibit
only a mild $\Gamma$ dependence such that similar
overall gains can still be achieved with just one
parameter set.

\begin{figure}
\centerline{\rotatebox{270}{\includegraphics[height=.49\textwidth,clip]{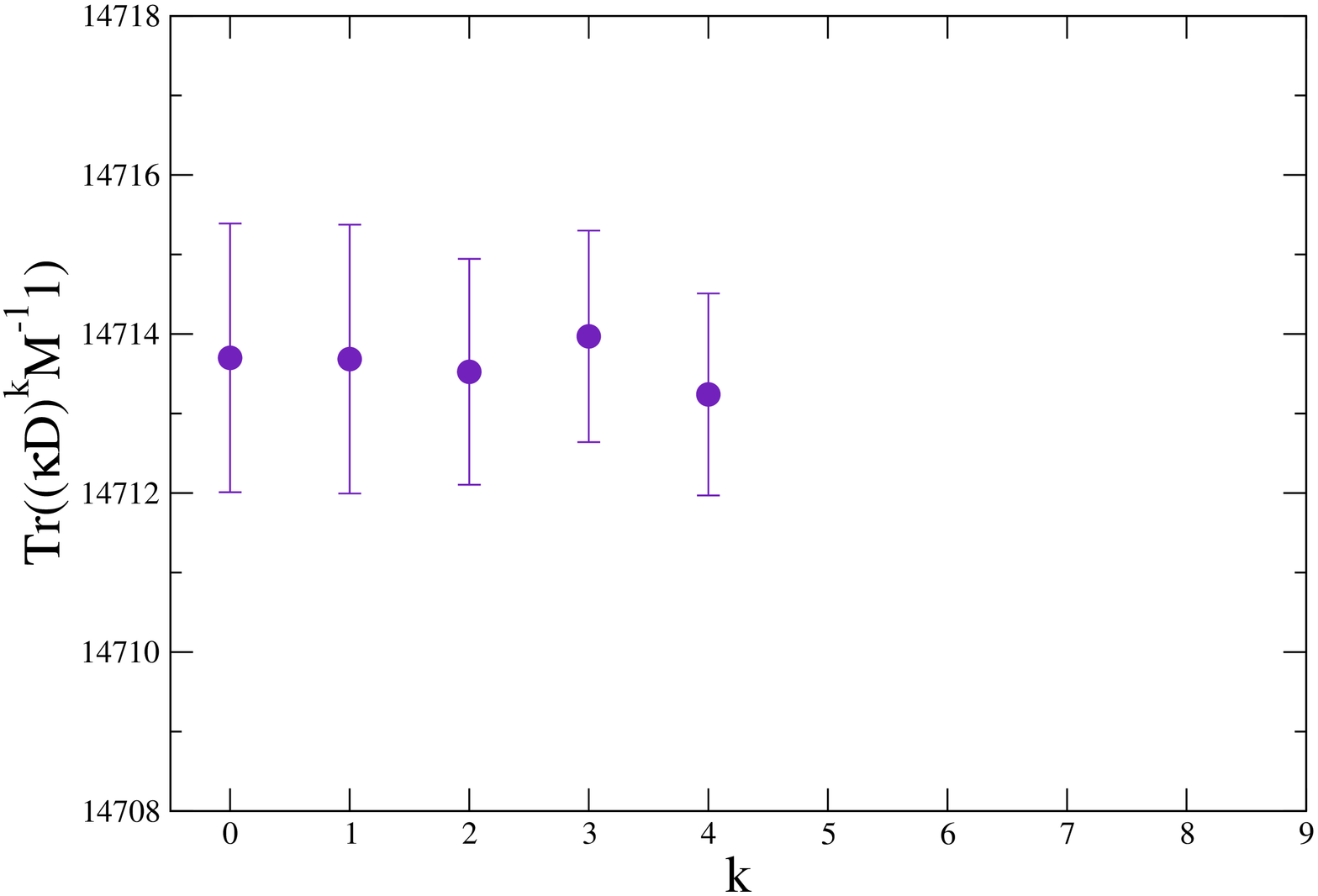}}\hspace{.02\textwidth}\rotatebox{270}{\includegraphics[height=.48\textwidth,clip]{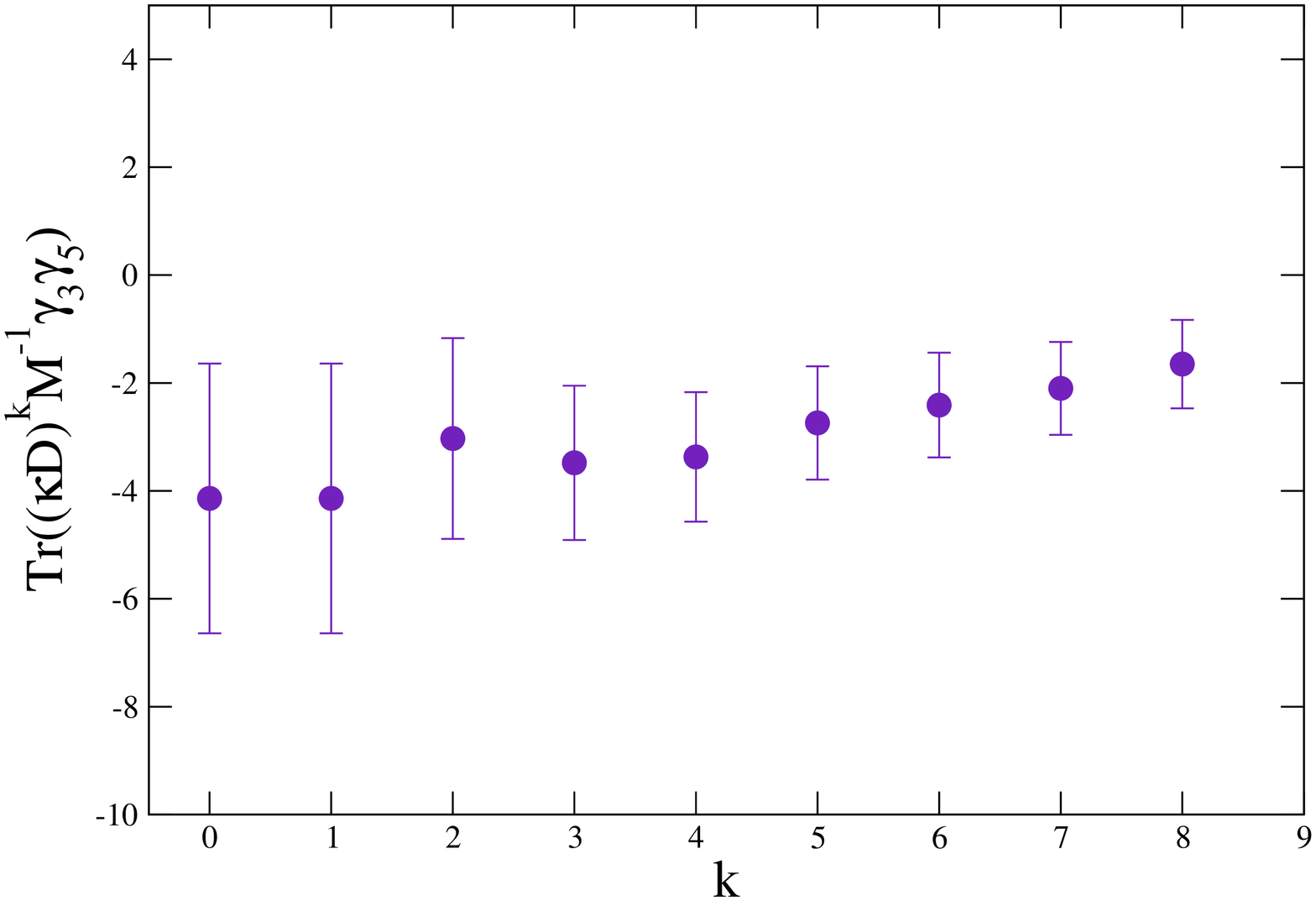}}}
\caption{Estimates of the zero momentum projected
$\mathrm{Tr}\,[(\kappa D)^kM^{-1}\Gamma]$
  at
  $\kappa_{\rm loop}=0.166$ for $\Gamma = \mathbb{1}$ (left)
and  $\Gamma=\gamma_3\gamma_5$ (right).
The errors are obtained from 300 estimates on one gauge configuration
and the zero-order HPE contribution for $\Gamma=\mathbb{1}$ was
calculated explicitly.
\label{hopping}}
\end{figure}

TSM can trivially be combined with the hopping parameter
expansion. At $\kappa_{\rm loop}=0.166$
the gains from applying a stand-alone HPE are almost as big as those from
the TSM. However, the HPE does not work well
for $\Gamma=\mathbb{1}$, see Fig.~\ref{hopping}.
Both methods
aim at removing noisy short range contributions from the
estimates. Clearly, the gain from combining the
two methods will be
smaller than the product of the two isolated gains.
However,  the separation of long and short range
is organized differently.
The HPE only works for ultra-local actions
and explicitly removes terms up to a particular lattice hopping radius.
It will be less convergent and hence less effective
at very light quark masses. The TSM is more generally
applicable and separates
short range modes from long range ones, but
this is done within the Krylov space of the solver.
When combining HPE with TSM, the reduced variances of the
truncated estimates result in larger optimal $n_{\rm t}$ values.
The constant computational overhead per solve
of the additional $D$ applications also
pushes $n_{\rm t}$ towards larger values.
Table~\ref{table:optvalues1} demonstrates that the
combined gain of these two methods
at $\kappa_{\rm loop}=0.166$ can be as large as a factor
of $30$.

\begin{table}
\begin{center}
\begin{tabular}{l|rrrrr|rrrrr}
\hline gain &
  \multicolumn{5}{c|}{TSM}& \multicolumn{5}{c}{TSM+HPE}\\\hline
  $m_{\rm PS}$ & $\mathbb{1}$ & $\gamma_3$ & $\gamma_1\gamma_2$ &
  $\gamma_5$ & $\gamma_3\gamma_5$& $\mathbb{1}$ & $\gamma_3$ &
  $\gamma_1\gamma_2$ & $\gamma_5$ & $\gamma_3\gamma_5$\\\hline
  600~MeV  & {$5$} &5 &10 & {$8$} & 8 & {$8$}& 11 & 19 & 25 &{$30$} \\
  450~MeV  & {$5$} &5 &10&  {$8$} & 8  & {$7$}&11 & 17 & 22 &{$25$} \\
  300~MeV & {$5$} &5 &10 & {$8$} & 8 & {$6$}& 9 & 15 & 17 & {$19$}
  \\\hline\end{tabular}
\end{center}
\caption{The TSM gains without and in
combination with the HPE at different quark masses.
  \label{table:massdep}}\end{table}

Having tested the method at $\kappa_{\rm loop}=0.166$, where
$m_{\rm PS}\approx 600$~MeV, we also use it to calculate
the disconnected loop at
$\kappa_{\rm loop}=0.1675$ and at $\kappa_{\rm loop}=0.1684$,
corresponding to pseudoscalars masses of approximately $450$~MeV and
$300$~MeV, respectively. The resulting gains are displayed
in Table~\ref{table:massdep}.
While the TSM performance is very  independent of the quark mass,
the HPE becomes less effective at lighter masses, in agreement
with our expectation. Still, even at the lightest quark mass,
the combined gains range from factors of
6 (for $\Gamma=\mathbb{1}$) up to 19 (for $\Gamma=\gamma_3\gamma_5$).

\begin{figure}
\centerline{\rotatebox{270}{\includegraphics[height=.9\textwidth,clip]{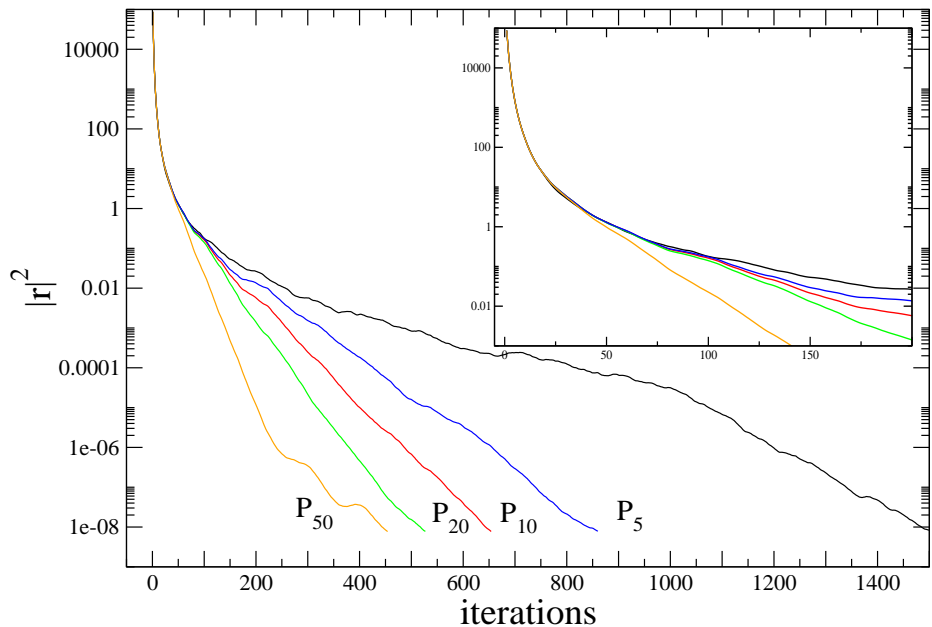}}}
\caption{Squared residual of the solver, as a function of the number of CG
iterations. $\mathbb{P}_{n_{\rm ev}}$ denote the outcomes, after deflating the
$n_{\rm ev}$ lowest modes.\label{fig:residual}}
\end{figure}

At light quark masses where the HPE becomes less effective,
the low eigenmode contributions to hadronic observables
might become more
dominant~\cite{Neff:2001zr,DeGrand:2004qw,Giusti:2004yp,Bali:2005pb}.
To study this effect, we deflate the lowest eigenmodes
at the lightest pseudoscalar mass.
As expected, this accelerates the
solver~\cite{deForcrand:1995bs,DeGrand:2002gm,Morgan:2004zh,Luscher:2007se,Stathopoulos:2007zi,Brannick:2007ue,Wilcox:2007ei,Darnell:2007dr,Tadano:2009gg}. We display a typical residual, as a function
of the number of CG iterations, without deflation and deflating the
lowest 5, 10, 20 and 50 modes of the Hermitian Wilson-Dirac operator in
Fig.~\ref{fig:residual}.

\begin{table}
\begin{center}
\begin{tabular}{l|rrrrr|rrrrr}
\hline
  $\mathrm{E}_{\rm imp}[\mathrm{Tr}\,(\mathrm{M}^{-1}\Gamma)]$ &\multicolumn{5}{c|}{TSM+HPE} &\multicolumn{5}{c}{TSM+HPE+TEA($\mathbb{P}_{20}$)}\\\hline
$\Gamma$ & $\mathbb{1}$ & $\gamma_3$ & $\gamma_1\gamma_2$ & $\gamma_5$ & $\gamma_3\gamma_5$& $\mathbb{1}$ & $\gamma_3$ & $\gamma_1\gamma_2$ & $\gamma_5$ & $\gamma_3\gamma_5$\\\hline
$n_{\rm t}$ & 155 & 220 & 155  & 160 & 240& 90 & 100 & 60  & 90 & 100\\
$N_1/N_2$ & 18 & 17 & 19 & 16 & 16 & 30 & 28 & 27 & 32 & 28 \\
$k$ &  4  & 4  & 4  & 8 & 8     &  4  &  4  & 4   & 8   &8\\\hline
gain (300)& 6 & 9 & 15 & 17 & 19&  13 & 21 & 30 & 38 & 49\\
gain (100)& 6  &10  & 16 &16  &20 &  5  &7  & 11 & 14 &20 \\\hline
\end{tabular}\end{center}
\caption{The optimized TSM parameters and
gains obtained at $\kappa_{\rm loop}=0.1684$, combining TSM with HPE,
with and without TEA (deflating 20 eigenmodes).
The gain factors are normalized to the costs of generating 300 and 100
standard estimates, respectively.
 \label{table:optvalues}}\end{table}

The optimized TSM parameters at $\kappa_{\rm loop}=0.1684$ for 
combining
TSM with HPE as well as for combining TSM with HPE and the TEA
of the lowest 20 eigenmodes are shown in Table~\ref{table:optvalues}.
The faster rate of convergence of the deflated solver leads to
smaller $n_{\rm t}$ values and therefore to larger $N_1/N_2$ ratios when
TSM is combined with TEA.
When normalized to the real cost of generating 300 unimproved stochastic
estimates, all gain factors increase by more than a factor of {\em two},
and this in spite of one quarter of the time being spent on generating
and projecting onto the eigenvectors. However, this factor
can fully be attributed
to the acceleration of the solver: while there exist observables
where stochastic errors decrease when applying the
TEA~\cite{Neff:2001zr,DeGrand:2004qw,Bali:2005pb}, for the quark loops that we
investigate here
this is not the case, at least not at pseudoscalar masses
above 300~MeV.
We observe a break-even between TSM+HPE+TEA(${\mathbb P}_{20}$)
and TSM+HPE when matching the cost of approximately 100 estimates, as can be
seen in the last
row of the table. In lattice simulations we will
encounter gauge errors from the Monte Carlo time series,
in addition to the errors from the stochastic estimates on single
configurations, discussed here. This interplay is studied
below.
\begin{figure}
\centerline{\rotatebox{270}{\includegraphics[height=.485\textwidth,clip]{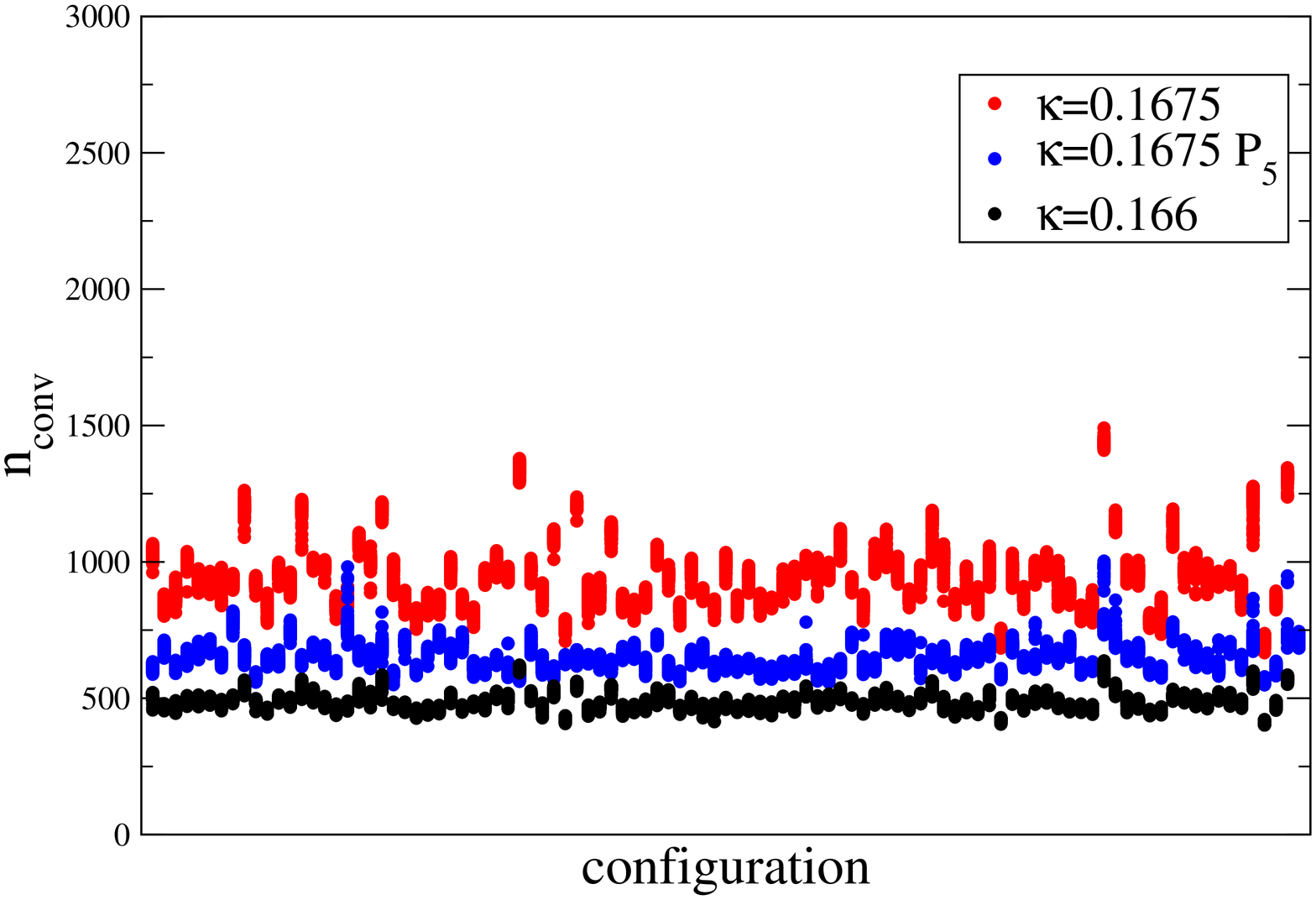}}\hspace{.02\textwidth}\rotatebox{270}{\includegraphics[height=.485\textwidth,clip]{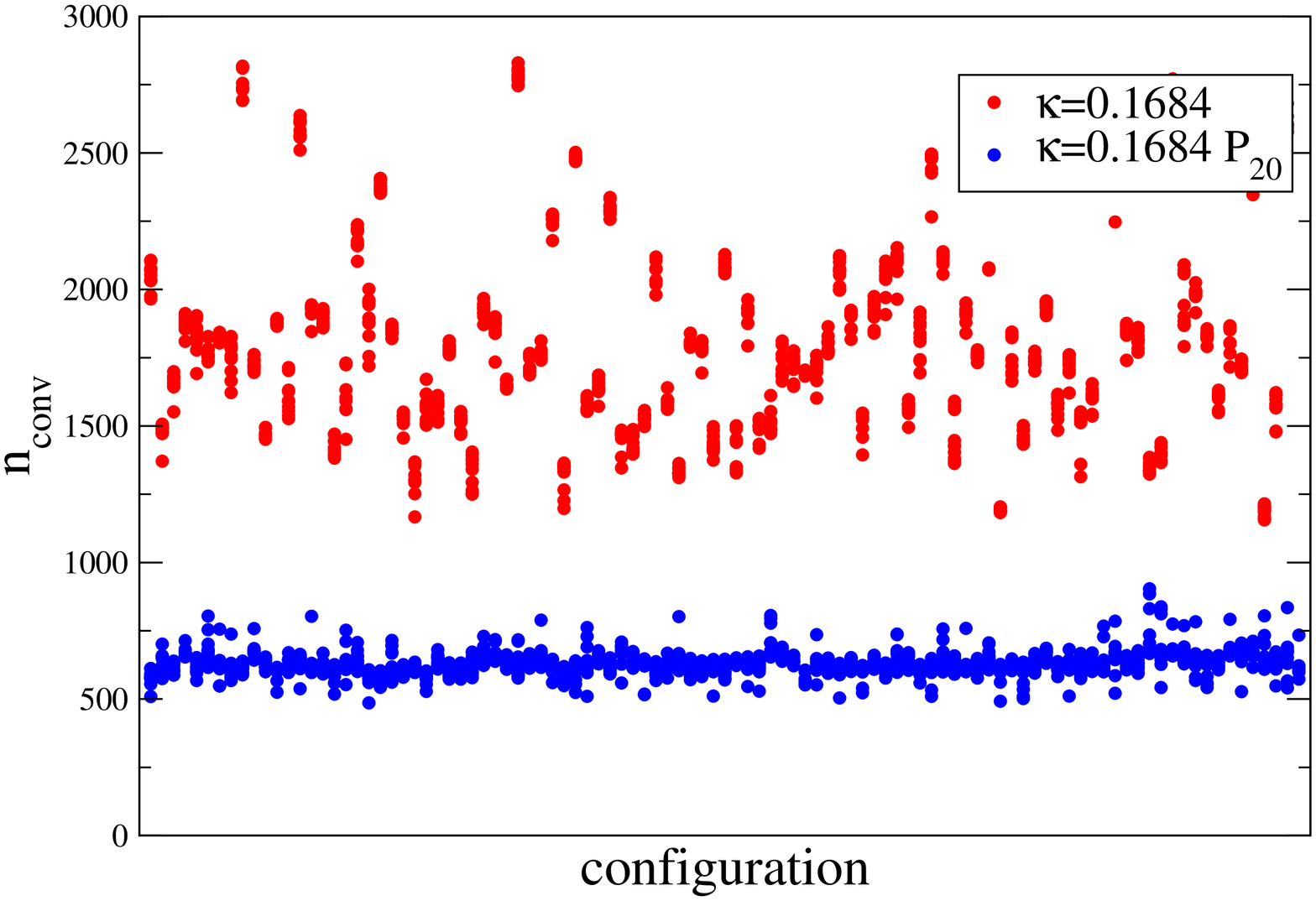}}}
\caption{The numbers of iterations to convergence. 
For the heavier two $\kappa$ values (left) the comparison is with 70
sources per configuration, for $\kappa=0.1684$ (right) we only utilize
10 sources.
\label{fig_opt}}
\end{figure}

\subsection{Configuration averages}
When calculating averages over configurations,
the stochastic errors $\sigma_{\rm stoch}$, that are defined in
Eq.~(\ref{eq:sto}), should be smaller than say half of the final
Monte Carlo errors $\sigma_{\rm tot}$. However, making
$\sigma_{\rm stoch}$ unnecessarily small can be a waste of
computer time that would be more wisely spent on analysing more
configurations or realizing additional source positions.
We start to study this balance of errors by computing
expectation values over configurations for the disconnected loops.  
In Sec.~\ref{sec:result} below we will then
calculate the three point functions that are of physical interest.

We remark that the (zero momentum projected)
gauge average $\langle\mathrm{Re}\,\mathrm{Tr}\,(M^{-1}{\mathbb 1})\rangle_c$
will
have a non-trivial value, while
$\langle\mathrm{Re}\,\mathrm{Tr}\,(M^{-1}\gamma_{\mu}\gamma_5)\rangle_c=0$.
We first assess the impact of low mode deflation.
In Fig.~\ref{fig_opt} we display the number of CG iterations
until convergence $n_{\rm conv}$ for several random sources on different
gauge configurations. Without deflation,
reducing the quark mass from $m_{\rm PS}\approx 600$~MeV
($\kappa=0.166$, black symbols of the left figure)
to $m_{\rm PS}\approx 300$~MeV
($\kappa=0.1684$, red symbols of the right figure), we find
$n_{\rm conv}$ to grow from about 500 to 1800 iterations.
This approximately four-fold increase is consistent
with the expected $1/m_{\rm PS}^2$ behaviour of the condition number
of $M$ (we solve for $Q^2=M^{\dagger}M$).
Deflating 5 eigenmodes ($\mathbb{P}_5$) at $\kappa=0.1675$ reduces the
required number of iterations by one third (red versus blue symbols of
the left figure) and deflating 20 modes at
$\kappa=0.1684$ results in a two third reduction
(red versus blue symbols of the right figure). We observe
variations of $n_{\rm conv}$ between and within configurations
that increase with decreasing
quark mass, as has been investigated quantitatively
in~\cite{DelDebbio:2005qa}.
Deflating the lowest modes vastly reduces the variance in
$n_{\rm conv}$. Clearly, the TSM parameter tuning
benefits from this stabilization.

\begin{table}
\begin{center}
\begin{tabular}{ccrcccc}
\hline
$\Gamma$&&cost &$\mathrm{E}_{\rm imp}$& $\sigma_{\rm stoch}^{\rm imp}$ &$\mathrm{E}_0$& $\sigma_{\rm stoch}$ \\\hline
$\mathbb{1}$
&$\kappa_{\rm loop}=0.166$  &300&14702.6 (7)&0.04&&\\
&$m_{\rm PS}\approx 600$~MeV&200&14702.6 (7)&0.05&&\\
                    &$n=326$&100&14702.6 (7)&0.06&14702.9 (7)&0.17\\
                    &       & 50&14702.6 (7)&0.09&14703.0 (8)&0.23\\
                    &       & 25&14702.5 (7)&0.13&14703.1 (8)&0.33\\
                    &       & 12&14702.5 (7)&0.18&14703.5 (9)&0.47\\
                    &       &  6&14702.3 (8)&0.23&14703.7(1.0)&0.65\\\hline
&$\kappa_{\rm loop}=0.1675$ &300&14743.1(1.1)&0.06&&\\
&$m_{\rm PS}\approx 450$~MeV&200&14743.0(1.1)&0.08&&\\
                    &$n=167$&100&14743.2(1.1)&0.11&14743.3(1.1)&0.25\\
                    &       & 50&14743.2(1.1)&0.16&14743.8(1.1)&0.33\\
                    &       & 25&14743.2(1.1)&0.23&14744.2(1.2)&0.47\\
                    &       & 12&14743.4(1.2)&0.33&14745.0(1.3)&0.69\\
                    &       &  6&14743.5(1.2)&0.42&14744.6(1.5)&0.96\\\hline
&$\kappa_{\rm loop}=0.1684$ &300&14764.9(1.2)&0.04&&\\
&$m_{\rm PS}\approx 300$~MeV&200&14764.9(1.2)&0.05&&\\
                    &$n=152$&100&14764.9(1.2)&0.08&14764.6(1.2)&0.27\\
                    &       & 90&14765.0(1.2)&0.10&14764.6(1.2)&0.29\\
                    &    &50$^*$&14765.0(1.2)&0.13&&\\
                    &    &25$^*$&14764.9(1.2)&0.19&&\\\hline\hline
$\frac{1}{3}\sum_j\gamma_j\gamma_5$
&$\kappa_{\rm loop}=0.166$  &300&-0.008 (50)&0.016&&\\
&$m_{\rm PS}\approx 600$~MeV&200&~0.007 (51)&0.019&&\\
                    &$n=326$&100&-0.033 (55)&0.027&-0.185(148)&0.135\\
                          & & 50&-0.054 (64)&0.039&-0.446(201)& 0.186\\\hline
&$\kappa_{\rm loop}=0.1675$ &300&-0.085 (87)&0.030&&\\
&$m_{\rm PS}\approx 450$~MeV&200&-0.096 (91)&0.038&&\\
                    &$n=167$&100&-0.040(101)&0.054&~0.003(211)&0.198\\
                          & & 50&-0.038(114)&0.076&~0.056(265)&0.271\\\hline
&$\kappa_{\rm loop}=0.1684$ &300&-0.069 (95)&0.015&&\\
&$m_{\rm PS}\approx 300$~MeV&200&-0.067 (96)&0.020&&\\
                    &$n=152$&100&-0.068 (96)&0.036&-0.089(216)&0.212\\
                    &       & 90&-0.072 (99)&0.042&-0.042(227)&0.223\\
                 &       &50$^*$&-0.141(106)&0.057&&\\
                 &       &25$^*$&-0.199(120)&0.082&&\\\hline
\end{tabular}\end{center}
\caption{Monte Carlo and stochastic errors for the (TSM+HPE) improved
and unimproved estimates of $\mathrm{Re}\,\mathrm{Tr}\,(M^{-1}\Gamma)$. The stochastic errors have
been normalized according to Eq.~(\protect\ref{eq:sto}).
For $\kappa_{\rm loop}=0.1684$, 20 eigenmodes
were deflated (TEA). The cost is in terms of the number of
unimproved estimates. Within the asterisked rows, the
calculation of these eigenvectors is not folded into the
cost calculation.\label{tab:aver}}\end{table}

We now average our improved estimates over $n$ configurations,
where we employ TSM, HPE and, at the lightest $\kappa$ value,
TEA with a projection onto the 20 lowest modes. In addition,
we calculate up to 100 unimproved estimates on each configuration. 
The results as functions of the real computer time spent, in units
of the cost of one unimproved estimate, i.e.\ of one solve,
are displayed in Table~\ref{tab:aver},
for $\Gamma=\mathbb{1}$ and
$\Gamma=\gamma_j\gamma_5$, where we average over the three
possible $j$-directions.
The total errors $\sigma_{\rm tot}$ are displayed
in brackets, followed by their respective lower bounds, as given by the
stochastic errors $\sigma_{\rm stoch}$.

For $\Gamma=\mathbb{1}$, even at the cost equivalent of six
standard estimates, the gauge errors still dominate over the
improved stochastic errors. Therefore, there is little point
in exceeding this number. On the given ensemble,
the same error can be obtained using 25 to 50 standard estimates,
suggesting a five-fold saving in computer time. However,
this is somewhat misleading since, by just increasing the number
of gauge configurations by a
factor of 1.6, the same error can be obtained, employing
six unimproved estimates.
It should be noted that the error balance could in principle look differently,
once the loop is correlated with a nucleon propagator, within a three
point function. Moreover, the stochastic error will become more
relevant on large volumes.
At the lightest $\kappa$ value, due to the overhead of setting
up TEA, the cost of the
improved estimates will always exceed the number 6. 
In fact, as can also be seen from Table~\ref{tab:aver},
TEA only turns out to be a worthwhile enterprise for a
cost equivalent bigger than 90.
Hence, unless the eigenvectors have been generated anyway, for instance
to enable low mode averaging~\cite{DeGrand:2004qw}
for the two point functions or for
the calculation of other current insertions, 
TEA is best omitted for $\Gamma=\mathbb{1}$.

The picture for the axial current containing
$\Gamma=\gamma_j\gamma_5$ is different: here at
the cost equivalent of 100 standard estimates, the stochastic error
still accounts for one quarter of the total error and
the gain of applying the improved method is in all cases
more than four-fold. This advantage should increase further
at larger volumes. Note that the reductions of the stochastic
errors alone, for the scalar and axial cases, are consistent
with the results of Table~\ref{table:massdep} that were
obtained on a single configuration.

\begin{figure}
\centerline{\rotatebox{270}{\includegraphics[height=.497\textwidth,clip]{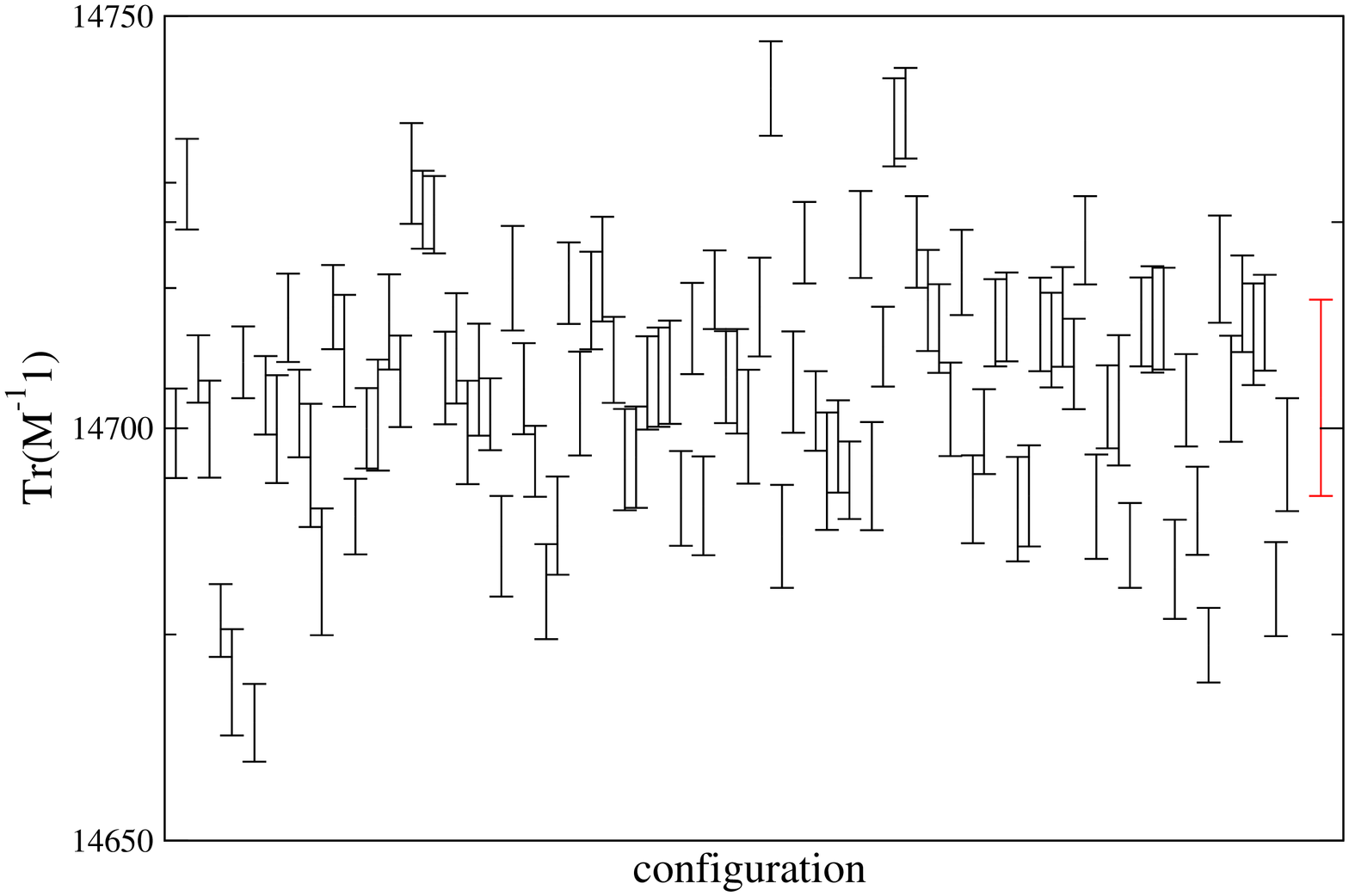}}\hspace{.02\textwidth}\rotatebox{270}{\includegraphics[height=.480\textwidth,clip]{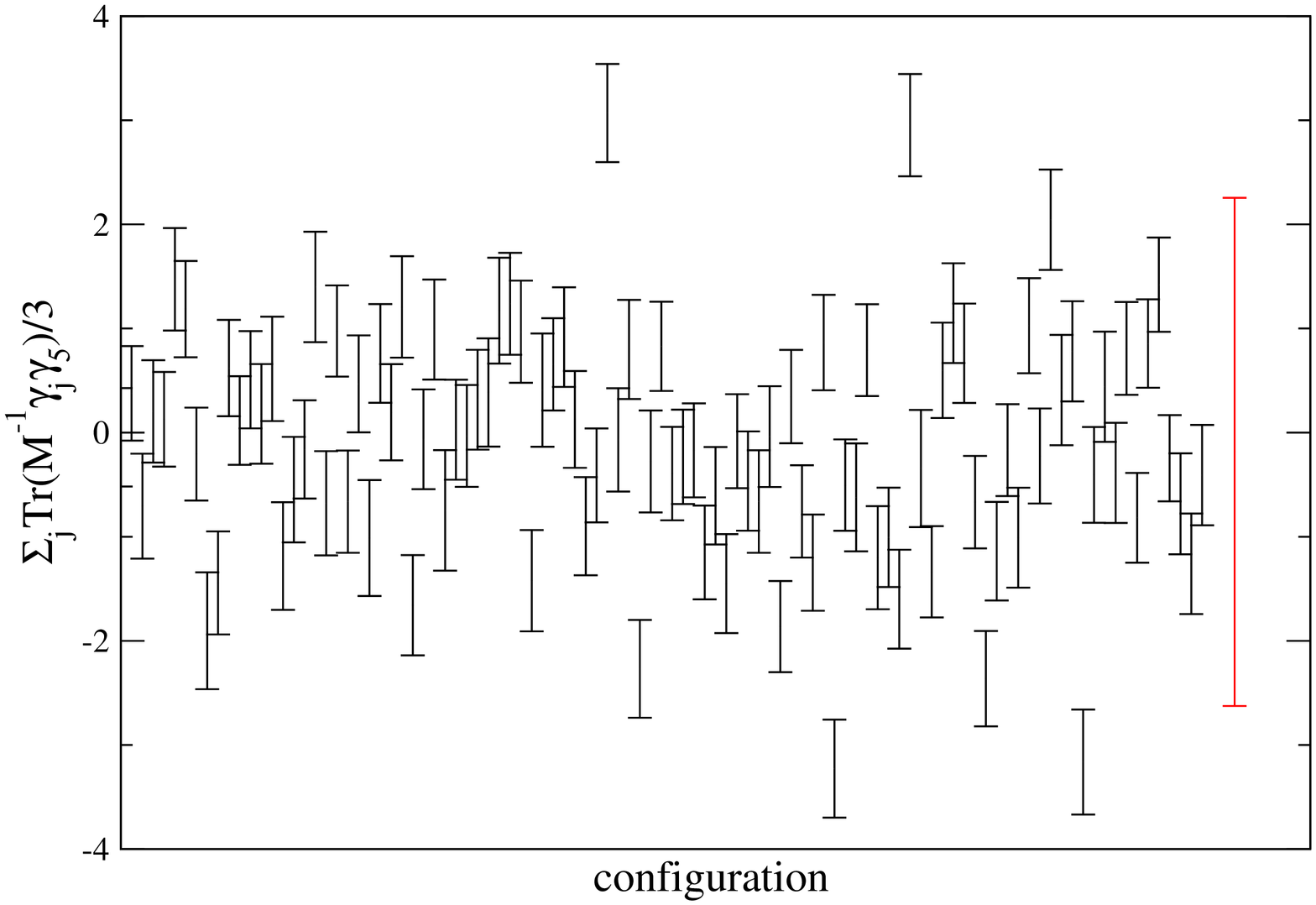}}}
\caption{The zero momentum projected estimates
$\mathrm{E}_{\rm imp}[\mathrm{Re}\,\mathrm{Tr}\,(M^{-1}{\mathbb 1})]$ (left)
and
$\frac13\sum_j
\mathrm{E}_{\rm imp}[\mathrm{Re}\,\mathrm{Tr}\,(M^{-1}\gamma_j\gamma_5)]$ (right)
at $\kappa=0.166$,
with stochastic errors $\Delta_{\rm stoch}$, on different
gauge configurations. The scalar loop was generated investing the
cost equivalent of 6 CG solves. The axial loop was evaluated at the
cost of 100 solves. The large error bars on the right of the figures correspond
to unimproved estimates $\mathrm{E}_0$.\label{loop_cfgs}}
\end{figure}

Based on these results, we decide to invest the cost
equivalent of 100 even/odd preconditioned
CG solves into the TSM/HPE/(TEA) estimation
of the axial loop and of 6 CG solves into the estimation
of the scalar loop. Note that for the calculation of a standard
point-to-all baryonic two-point correlation function usually 12
such solves are
required, and even more sources if a variational basis is used
to optimize the creation operator. We display the resulting
stochastic errors $\Delta_{\rm stoch}$ on single configurations 
at $\kappa_{\rm loop}=0.166$
in Fig.~\ref{loop_cfgs}. 
This graphical representation again makes
it evident that even for this low-cost setting the stochastic errors are 
much smaller than the gauge noise.
The larger error bars on the right of the figures correspond to
unimproved $\Delta_{\rm stoch}$ values, obtained at the
same cost. After application of
the improvement methods, the stochastic errors are so small that nothing
extra can be gained from increasing the number of stochastic sources
(and solves) beyond these moderate values.

\section{Application to $\Delta q^{\rm dis}$ and to
$\langle N|\bar{q}q|N\rangle^{\rm dis}$}
\label{sec:result}
\subsection{Definition of the matrix elements}
We now apply our methods to the calculation
of observables of phenomenological interest, namely
of $\Delta q$ and $\langle N|\bar{q}q|N\rangle$. 
The contribution to the nucleon spin $\Delta q$
is defined through the matrix element (in Minkowski space
notation),
\begin{equation}
\langle N,s|\bar{q}\gamma_{\mu}\gamma_5 q|N,s\rangle = 2m_N s_{\mu}
\frac{\Delta q}{2}\,,
\end{equation}
where $m_N$ denotes the mass of the nucleon $N$ and
$s_{\mu}$ its spin ($s_\mu^2=-1$).
$\Delta q$
and 
$\langle N|\bar{q}q|N\rangle$
are extracted from the ratios of
three-point functions to two-point functions~(at zero momentum):
\begin{equation}
\label{eq:rati}
R^{\rm dis}(t,t_{\rm f}) = 
-
\frac{\mathrm{Re}\,\left\langle\Gamma_{\rm 2pt}^{\alpha\beta}C^{\beta\alpha}_{\rm 2pt}(t_{\rm f}) \sum_{\mathbf{x}}\mathrm{Tr}\,(M^{-1}(\mathbf{x},t;\mathbf{x},t)\Gamma_{\rm loop})\right\rangle_c}{\left\langle \Gamma_{\rm unpol}^{\alpha\beta} C^{\beta\alpha}_{\rm 2pt}(t_{\rm f})\right\rangle_c}\,.
\end{equation}
For the scalar matrix element we use,
$\Gamma_{\rm 2pt}=\Gamma_{\rm unpol}:=(1+\gamma_4)/2$ and
$\Gamma_{\rm loop}=\mathbb{1}$. For
$\Delta q$ we calculate the difference between
two polarizations:
$\Gamma_{\rm 2pt}= \gamma_j\gamma_5\Gamma_{\rm unpol}$ and
$\Gamma_{\rm loop}=\gamma_j\gamma_5$, where we average over
all three possible $j$-orientations. The spin projection
operators along the $j$-axis
read, $P_{\uparrow\downarrow}=\frac12(\mathbb{1}\pm i\gamma_j\gamma_5)$,
so that in this case,
$\Gamma_{\rm 2pt}= -i(P_{\uparrow}-P_{\downarrow})\Gamma_{\rm unpol}$,
where we have traded a factor $i$ against taking the real part,
rather than the imaginary part, of the nominator
in Eq.~(\ref{eq:rati}).
The variance of the above expression is reduced by
explicitly using the fact that $\mathrm{Im}\,\mathrm{Tr}\,(M^{-1}{\mathbb 1})
=\mathrm{Im}\,\mathrm{Tr}\,(M^{-1}\gamma_j\gamma_5)=0$.

The two-point function 
of the zero momentum projected proton with sink and source
spinor indices $\alpha$ and $\beta$
is given by,
\begin{equation}
C^{\alpha\beta}_{\rm 2pt}(t_{\rm f})=
\sum_{\mathbf x}
\left\langle 0\left|
N^{\alpha}({\mathbf x},t_{\rm f}){N^{\beta}}^{\dagger}({\mathbf 0},0)
\right| 0\right\rangle\,,
\end{equation}
where we have set $t_0=0$.
This can be constructed from standard point-to-all
quark propagators~\cite{DeGrand:2006zz}.
Note that for
$q=u,d$ there are additional connected contributions $R^{\rm con}$, which
we have not calculated. We combine the three $\kappa_{\rm loop}$ values
with $\kappa_{\rm 2pt}=0.166$ and 0.1675.

In the limit of large times, $t_{\rm f}\gg t\gg 0$, in the axial case,
\begin{equation}
\label{eq:tlimit}
R^{\rm dis}(t,t_{\rm f})+R^{\rm con}(t,t_{\rm f}) \rightarrow
2\frac{\langle N,s|\bar{q}\gamma_{j}\gamma_5q|N,s\rangle}{2m_N}
=\Delta q\,.
\end{equation}
The prefactor {\em two} comes from taking the difference between
two opposite polarizations rather than fixing one polarization. 
We note that this result will be in
a lattice scheme and still needs to be multiplied by a renormalization
factor of $\mathcal{O}(1)$, for a translation into the $\overline{MS}$ scheme.
In the scalar case, 
$\langle N|\bar{q}q|N\rangle$ is defined as the connected contribution
only and can thus be obtained by subtracting
$\langle 0|\bar{q}q|0\rangle=
-\sum_{\mathbf{x}}\mathrm{Re}\left\langle\mathrm{Tr}\,(M^{-1}(\mathbf{x},t;\mathbf{x},t)\right\rangle_c$ from Eq.~(\ref{eq:rati}).
We will label the disconnected contributions to these two matrix elements
as, $\Delta q^{\rm dis}$ and $\langle N|\bar{q}q|N\rangle^{\rm dis}$,
respectively. In the case of the strange quark,
$\Delta s=\Delta s^{\rm dis}$ and $\langle N|\bar{s}s|N\rangle=
\langle N|\bar{s}s|N\rangle^{\rm dis}$.
\begin{figure}
\centerline{\rotatebox{270}{\includegraphics[height=.9\textwidth,clip]{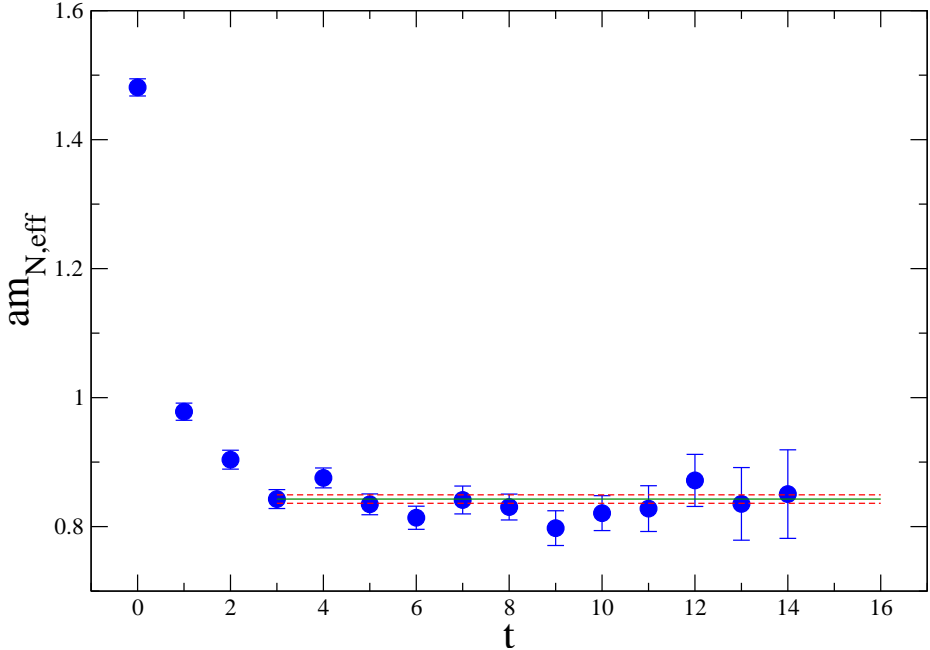}}}
\caption{The effective mass, Eq.~(\protect\ref{eq:effma}),
of the nucleon smeared at source and
sink for $\kappa=0.166$. The time $t$ is displayed
in lattice units $a\approx 0.13$~fm.
The result of a fit to the time range $3\leq t< 16$ is shown
as horizontal lines.\label{smearing}}
\end{figure}

\subsection{Results of the calculation}

We calculate the disconnected loop on only one timeslice $t=3\approx 0.38\,
\mathrm{fm}/a$. Having the operator inserted close to
the source ($t_0=0$) reduces the statistical errors
but care must be taken that
contributions from excited states are suppressed so that
Eq.~(\ref{eq:tlimit}) still holds. Following \cite{Bali:2005pb},
we use Wuppertal smearing on top of APE
smeared links at the source and sink for both the three point and two
point functions. In Fig.~\ref{smearing} we display
the effective mass,
\begin{equation}
\label{eq:effma}
m_{N,\mathrm{eff}}(t)=a^{-1}\ln\left(\frac{C_{\rm 2pt}(t)}{C_{\rm 2pt}(t+1)}\right)\,,
\end{equation}
as a function of the time. This
shows that with our choice of
smearing parameters the excited state contributions are not
significant at $t=3$. Hence, in the symmetric situation,
$t_{\rm f}=6$, such contributions should be negligible.
This is even more so
since the statistical error will be much bigger than that of
the two point function at $t=3$. The consistency of this
assumption can be checked by varying $t_{\rm f}\geq 4$.

\begin{figure}
\centerline{\rotatebox{270}{\includegraphics[height=.480\textwidth,clip]{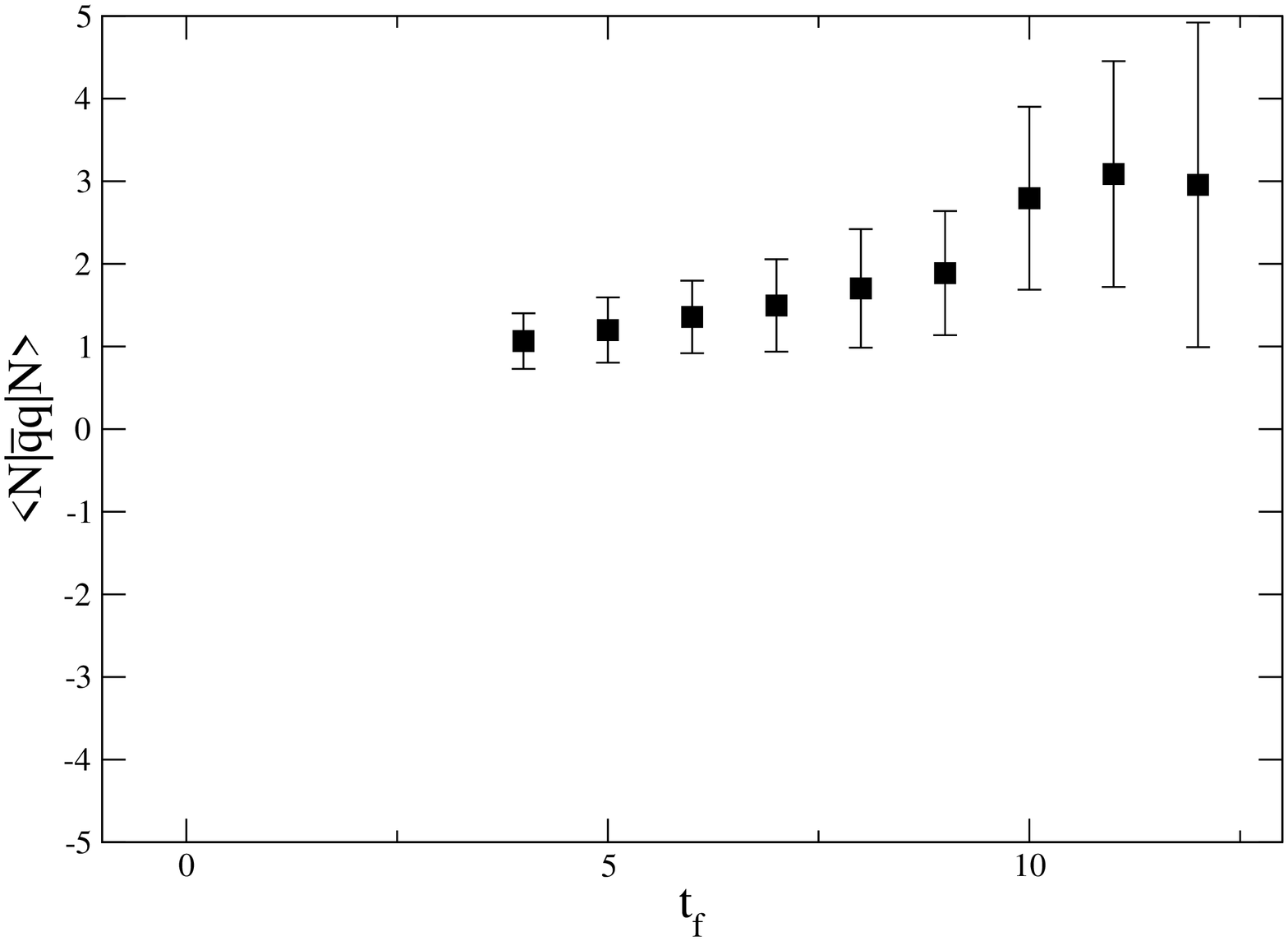}}\hspace{.02\textwidth}\rotatebox{270}{\includegraphics[height=.497\textwidth,clip]{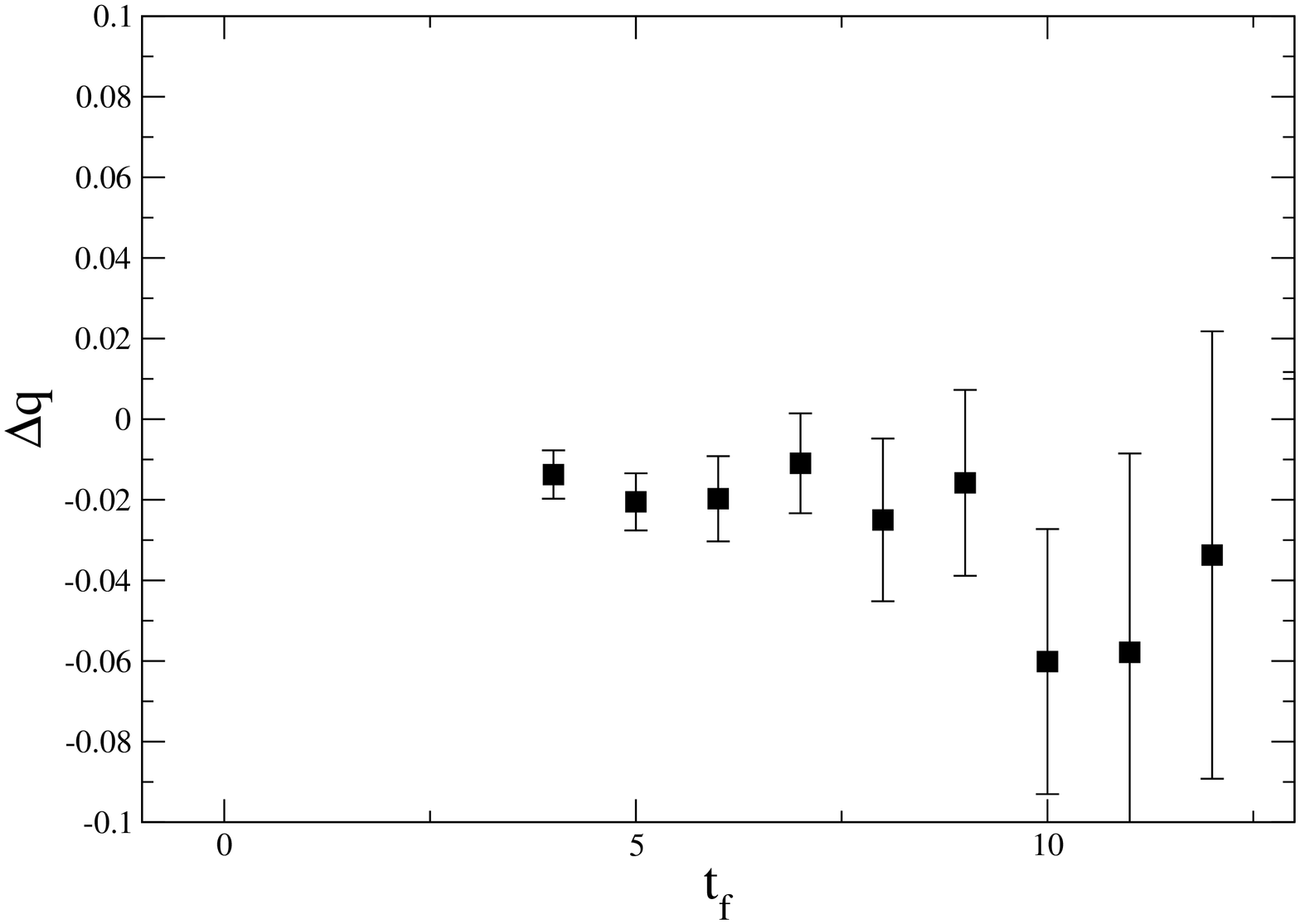}}}
\caption{The matrix elements $\langle N|\bar{q}q|N\rangle^{\rm dis}$ (left)
and $\Delta q^{\rm dis}$ (right) for different values of $t_{\rm f}$ at
$\kappa_{\rm loop}=\kappa_{\rm 2pt}=0.166$.\label{ratiodq}}
\end{figure}
In Fig.~\ref{ratiodq} we display finite time estimates of the
matrix elements $\langle N|\bar{q}q|N\rangle^{\rm dis}$ and
$\Delta q^{\rm dis}$ in the lattice scheme,
for different final times $t_{\rm f}$ at
$\kappa_{\rm loop}=\kappa_{\rm 2pt}=0.166$.
This value approximately corresponds to the mass of the strange quark.
The stochastic estimates of the loops were generated using time partitioning
($t=3$), TSM and HPE, at the cost of 6 even/odd preconditioned CG solves
for the scalar and at the cost of 100 such solves for the axial current.
Our analysis of the two point function above suggests that finite $t$
systematics should be small, relative to the statistical errors at $t_{\rm f}=6$.
Indeed, all $t_{\rm f}\geq 4$ data are consistent with a constant. We 
conservatively quote the $t_{\rm f}=6$ values as our final results.
We realized all six $\kappa_{\rm loop}\in\{0.166, 0.1675, 0.1684\}$
and $\kappa_{\rm 2pt}\in\{0.166, 0.1675\}$ combinations where 
the lightest $\kappa_{\rm loop}$ value corresponds to
$m_{\rm PS}\approx 300$~MeV. The five combinations that are not shown
result in the same
general picture, with larger statistical errors.
For the two point function we do not realize the lightest $\kappa$ value
since, without low mode averaging, this turns out to be very noisy.

\begin{table}
\begin{center}
\begin{tabular}{r|ll|ll}
\hline 
 \multicolumn{5}{c}{$\kappa_{\rm loop}=0.166$}\\\hline
 & \multicolumn{2}{c|}{$\kappa_{\rm 2pt}=0.166$} &
       \multicolumn{2}{c}{$\kappa_{\rm 2pt}=0.1675$}\\\hline
cost     &  $\langle N|\bar{q}q|N\rangle_{\rm imp}^{\rm dis}$ & $\langle N|\bar{q}q|N\rangle_0^{\rm dis}$  & $\langle N|\bar{q}q|N\rangle_{\rm imp}^{\rm dis}$ & $\langle N|\bar{q}q|N\rangle_0^{\rm dis}$\\\hline
300 & 1.57(43) &         & 1.90(54)   &   \\
200 & 1.58(43) &         & 1.91(54)   &  \\
100 & 1.62(43) & 1.54(45)& 1.95(53)   & 1.92(57)         \\
 50 & 1.65(42) & 1.68(46)& 2.00(54)   & 2.00(61) \\
 25 & 1.62(42) & 1.65(51)& 1.99(54)   & 2.15(71) \\
 12 & 1.48(44) & 1.85(51)& 1.93(59)   & 2.44(71)   \\
 6  & 1.36(44) & 1.79(62) &1.81(57)   & 2.19(79)\\\hline
 \multicolumn{5}{c}{$\kappa_{\rm loop}=0.1675$}\\\hline
 300 & 1.39(63)   &             & 1.36(69)   & \\
 200 & 1.38(63)   &             & 1.34(69)  &\\
 100 & 1.35(65)   &  1.33(66)   & 1.33(72)  & 1.16(72)\\
 50 &  1.47(67)   &  1.50(68)   & 1.44(74)  & 1.29(74)\\
 25 &  1.49(70)   &  1.62(74)   & 1.36(79)  & 1.40(81)\\
  12 & 1.33(73)   &  1.91(76)   & 1.24(85)  & 1.64(85)\\
   6 & 1.38(77)   &  1.72(88)   & 1.30(89)  & 1.28(99)\\\hline
  \multicolumn{5}{c}{$\kappa_{\rm loop}=0.1684$}\\\hline
300 &  1.45(73)     &         &  1.39(77)     &   \\
200 &  1.44(73)     &         &  1.38(78)     &   \\
100 &  1.44(74)     & 0.94(64)&  1.38(78)   &  0.88(76)      \\
 90 &  1.43(74)     & 0.92(63)&  1.36(78)   &  0.87(75)  \\
50$^*$&1.41(74)     &         &  1.30(78)   & \\
25$^*$&1.50(74)     &         &  1.40(78)   & \\
12$^*$&1.61(74)     &         &  1.51(77)   & \\
 6$^*$&1.60(74)     &         &  1.52(76)   & \\\hline
\end{tabular}
\end{center}
\caption{The disconnected contribution to the
scalar matrix element in the lattice scheme,
evaluated at different cost
equivalents. At $\kappa_{\rm loop}=0.1684$ TEA ($\mathbb{P}_{20}$) is
employed. In the asterisked rows the costs
of generating the eigenvectors
were neglected.\label{tab:scalar}}
\end{table}
\begin{figure}
\centerline{\rotatebox{270}{\includegraphics[height=.472\textwidth,clip]{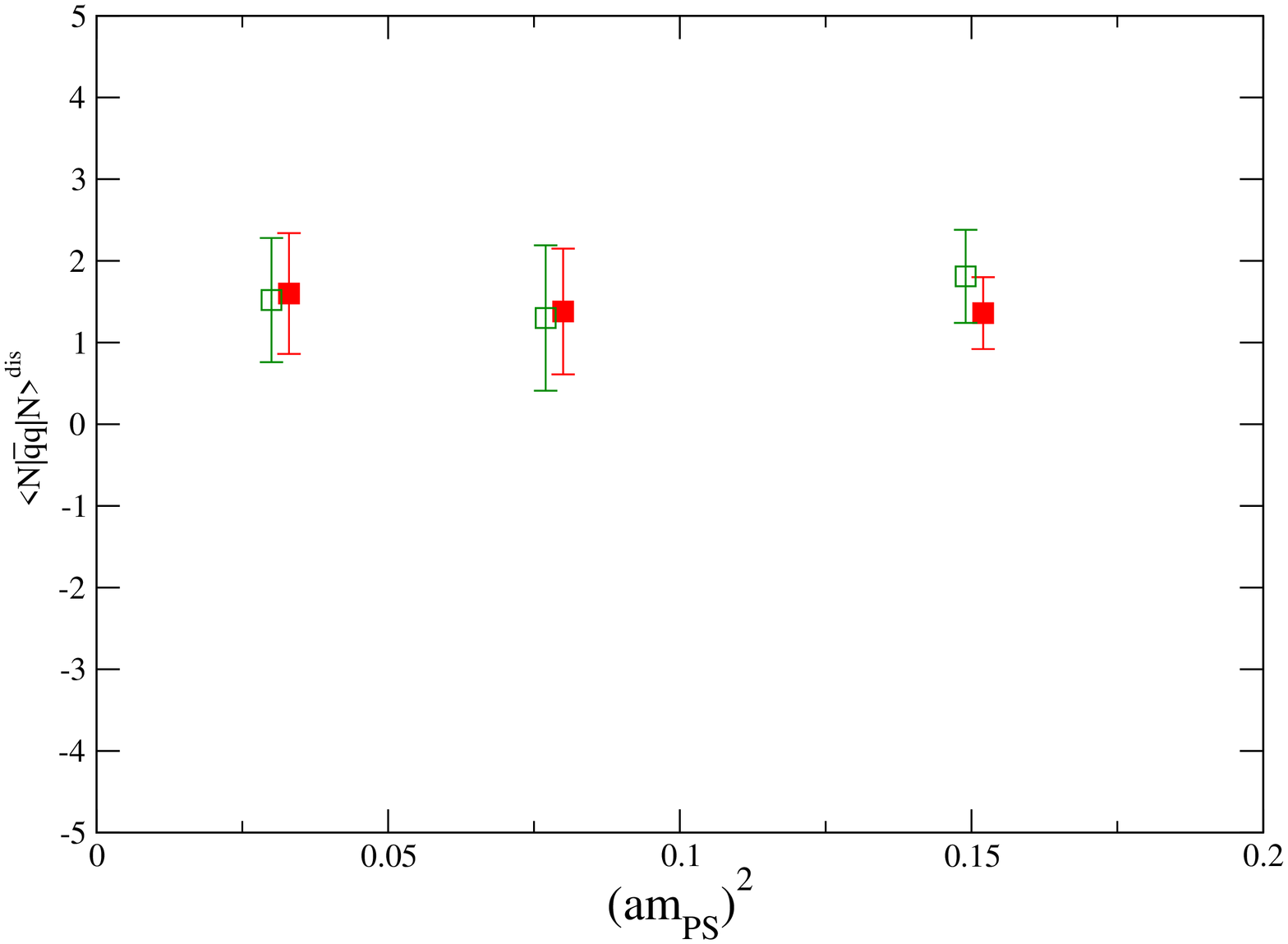}}\hspace{.02\textwidth}\rotatebox{270}{\includegraphics[height=.495\textwidth,clip]{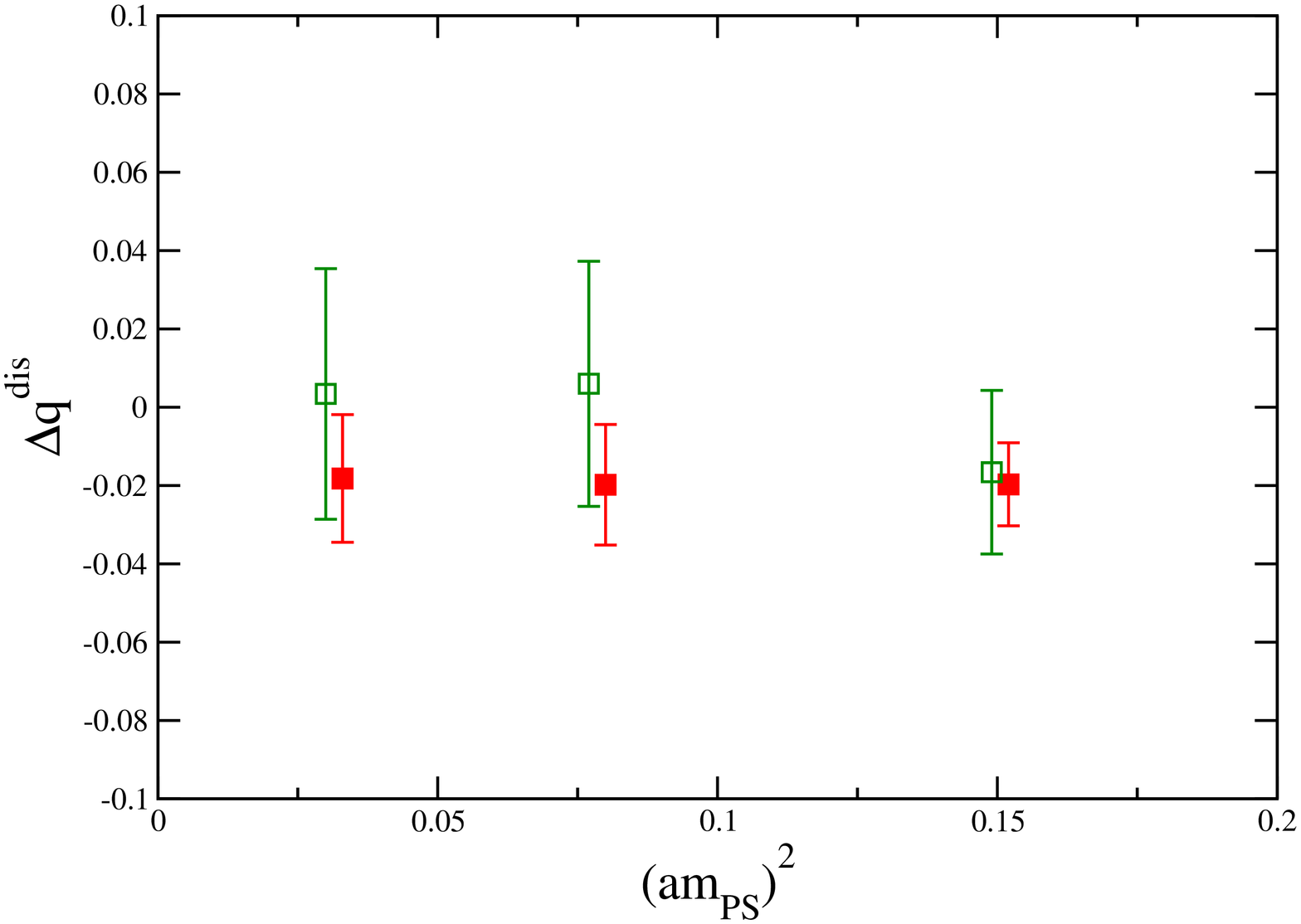}}}
\caption{$\langle N|\bar{q}q|N\rangle^{\rm dis}$~(left) and $\Delta
  q^{\rm dis}$~(right) as functions of the quark mass used in the
  disconnected loop~(expressed in terms of $am_{\rm PS}^2$).
 The open squares correspond to a proton with
  $\kappa_{\rm 2pt}=0.1675$, the filled squares
to the heavier $\kappa_{\rm 2pt}=0.166$.\label{extrap_delq}}
\end{figure}

In Table~\ref{tab:scalar} we display improved and conventional
stochastic estimates of the scalar matrix elements (obtained at
$t_0=0$, $t=3$ and $t_{\rm f}=6\approx 0.76\,\mathrm{fm}/a$), for all
our $\kappa$ combinations. The fixed cost reductions, due to
TSM and HPE, in the
relative errors are small and do by far not match the
gain factors that we obtained in
Sec.~\ref{sec:discon} for the disconnected loops alone.
The noise is dominated by taking the correlation with the
two point function.
For the precision
of the disconnected loop to be matched by that of the two point function,
the latter needs to be evaluated for multiple source points,
eventually in combination
with low mode averaging~\cite{DeGrand:2004qw}.
We note that one does not encounter any computational overhead
in calculating the disconnected loop at more than one timeslice.
As long as these are sufficiently separated, the stochastic
errors will not increase significantly.

\begin{table}
\begin{center}
\begin{tabular}{r|cc|cc}
\hline 
 \multicolumn{5}{c}{$\kappa_{\rm loop}=0.166$}\\\hline
 & \multicolumn{2}{c|}{$\kappa_{\rm 2pt}=0.166$} &
       \multicolumn{2}{c}{$\kappa_{\rm 2pt}=0.1675$}\\\hline
cost     & $\Delta q_{\rm imp}^{\rm dis}$ &$\Delta q_0^{\rm dis}$   &$\Delta q_{\rm imp}^{\rm dis}$&$\Delta q_0^{\rm dis}$\\\hline
300  &-0.025 (9) &            & -0.023(17)   & \\
200  &-0.026 (9) &            & -0.028(18)   &  \\
100  &-0.020(11) & -0.020(30) & -0.017(21)   & ~0.015(54) \\
 50  &-0.021(13) & ~0.001(39) & -0.024(26)   & ~0.052(73)\\\hline
  \multicolumn{5}{c}{$\kappa_{\rm loop}=0.1675$}\\\hline
300 & -0.026(14)  &          & -0.009(28)  &   \\
200 & -0.027(15)  &          & -0.017(28) & \\
100 & -0.020(15) & -0.006(34)& ~0.006(31) & ~0.067(70)\\
 50 & ~0.009(18) & ~0.067(39)& ~0.041(37) & ~0.194(92)\\\hline
  \multicolumn{5}{c}{$\kappa_{\rm loop}=0.1684$}\\\hline
300 &   -0.016(15)   &             &   ~0.010(31)   &   \\
200 &   -0.016(15)   &             &   ~0.010(30)   &    \\
100 &   -0.019(16)   & ~0.028(41)  &   ~0.002(31)   & ~0.017(67)\\
 90 &   -0.018(16)   & ~0.024(43)  &   ~0.003(32)   & ~0.011(65)  \\
50$^*$& -0.026(18)   &             &   -0.008(37)   & \\
25$^*$& -0.016(20)   &             &   ~0.011(38)   & \\\hline
\end{tabular}
\end{center}
\caption{Disconnected contribution to $\Delta q$ in the lattice scheme,
evaluated at different cost
equivalents. At $\kappa_{\rm loop}=0.1684$ TEA ($\mathbb{P}_{20}$) is
employed. In the asterisked rows the costs
of generating the eigenvectors
were neglected.\label{tab:axial}}
\end{table}

In Table~\ref{tab:axial} we display the same information as
in Table~\ref{tab:scalar}, for $\Delta q^{\rm dis}$ in the lattice scheme.
In this case, at the cost equivalent of 100 estimates (90
estimates at $\kappa_{\rm loop}=0.1684$),
the gains of applying TSM and HPE in terms of computer time,
relative to time partitioning alone, are about ten-fold.
The reductions in error are close to those displayed in
Table~\ref{tab:aver} for the disconnected loop alone.
This channel is not yet limited by the accuracy of the two point
function but of course also in this case statistics could be increased
by averaging over multiple baryon sources and over forward as well
as backward propagation.

In Fig.~\ref{extrap_delq} we display our final
$t_{\rm f}=6$ results for the two matrix elements.
In neither case do we
observe
any significant dependence on the valence quark mass, varying this from
$m_{\rm PS}\approx 600$~MeV down to 450~MeV, or on the 
loop quark mass, reducing $m_{\rm PS}\approx 600$~MeV
(strange quark mass) to $m_{\rm PS}\approx 300$~MeV.
We find 
$\Delta s=-0.020(11)$ at the heavier proton mass and
$\Delta s=-0.017(21)$ at the lighter mass value, 
while the scalar
matrix elements are somewhat larger than one.
Note that the lattice results presented here are
unrenormalized. However, based on perturbative
two loop results~\cite{Skouroupathis:2008mf}, albeit
with different sea quark and gluon actions,
we would not expect $\Delta s$ to
change by more than a factor of 0.7 after a translation
into the $\overline{MS}$ scheme.

\section{Conclusions and outlook}
A growing number of Lattice QCD applications requires all-to-all
propagators. These are most efficiently estimated stochastically.
We presented the novel truncated solver method (TSM) that typically
reduces the computer time to achieve a given stochastic variance
by an order of magnitude, without introducing a bias.
The gain factors of this method for different observables 
were demonstrated not to vary when changing the quark mass
by a factor of four ($600\,\mathrm{MeV}\geq m_{\rm PS}\geq
300\,\mathrm{MeV}$). The TSM is easy to implement and can be
used for any fermionic action. The
combination of TSM with different variance reduction techniques is
straight forward and this reduces the stochastic
variance even further. We studied in detail combinations of TSM,
partitioning~\cite{Bernardson:1993yg}, the hopping parameter
expansion~\cite{Thron:1997iy} and low eigenmode
deflation~\cite{DeGrand:2002gm,Bali:2005pb}.

In realistic Lattice QCD simulations there are
intrinsic errors from the Monte Carlo time series, in addition to the
errors introduced by the stochastic estimation of the inverse
of the fermionic matrix on individual configurations.
We studied the
interplay between these gauge and stochastic noises.
After reducing the stochastic variance by a combination
of methods, in our calculation of disconnected contributions
to the nucleon structure, the gauge errors became the dominant sources
of uncertainty. This means that we can
afford larger stochastic errors and for instance increase the
lattice volume, without having to increase the number of random sources.
For instance, on our 2~fm lattices, even at a cost of
only 6 CG solves, the stochastic error of the scalar matrix
element $\langle N|\bar{q}q|N\rangle^{\rm dis}$ is completely
over-shadowed by its gauge error:
in certain situations 
one can overdo the reduction of the stochastic noise.
In this particular case the total error can
more efficiently be reduced by increasing the number
of nucleon sources on each configuration than by increasing
the number of (improved) estimates. Also the determination
of $\Delta s$ will benefit from this.
At present we are pursuing such an approach.

Our  result on the strangeness contribution to the nucleon spin
$\Delta s\approx 0$ is in agreement with the other recent
direct calculation of this quantity~\cite{Babich:2009rq}.
However, we disagree with earlier, less precise studies that
employed a summation method over
$t$~\cite{Fukugita:1994fh,Dong:1995rx,Gusken:1999as}.
These suggested a value $\Delta s\approx -0.12$. In the case of the scalar
matrix element our errors are still too large to state a meaningful
number, in particular since chiral and infinite volume behaviours
need to be studied. The techniques developed here are used by us
in an ongoing study~\cite{inprep} at smaller lattice spacing, quark masses
and larger volumes with Sheikholeslami-Wohlert sea and valence quarks.

\section*{Acknowledgements}
We thank Zoltan Fodor and Kalman Szabo for providing us with the gauge
configurations and Andreas Frommer for discussions. Our computations
were performed on the Regensburg Linux cluster and the
IBM p6 cluster (JUMP), BlueGene/L (JUBL)
and BlueGene/P (JUGENE) of the J\"ulich Supercomputer Center.
Sara Collins  acknowledges support from the
Claussen-Simon-Foundation (Stifterverband f\"ur die Deutsche
Wissenschaft). This work was supported by the DFG
Sonderforschungsbereich/Transregio 55.
\newpage
\begin{figure}
\centerline{\rotatebox{270}{\includegraphics[height=.485\textwidth,clip]{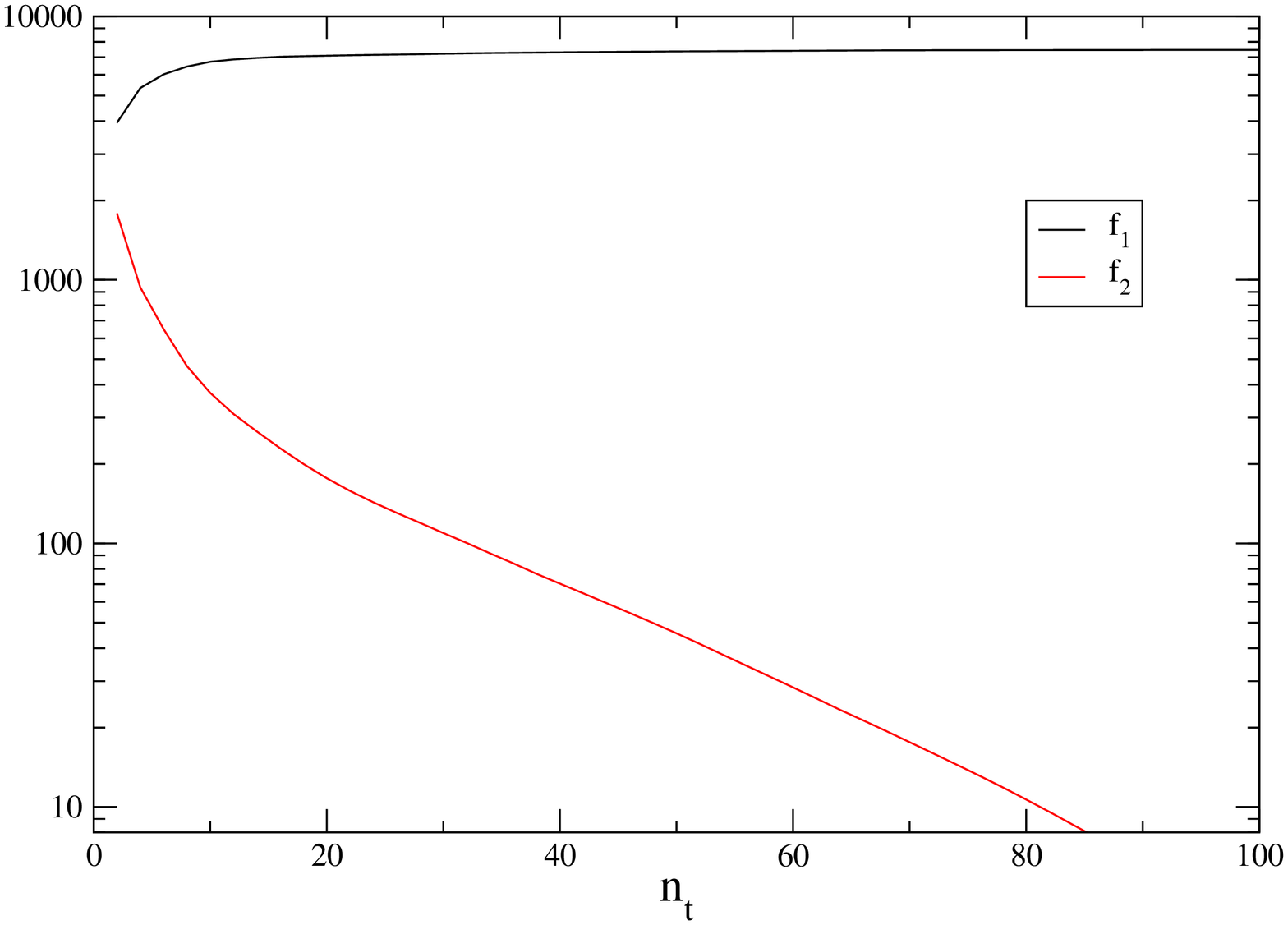}}\hspace{.02\textwidth}\rotatebox{270}{\includegraphics[height=.485\textwidth,clip]{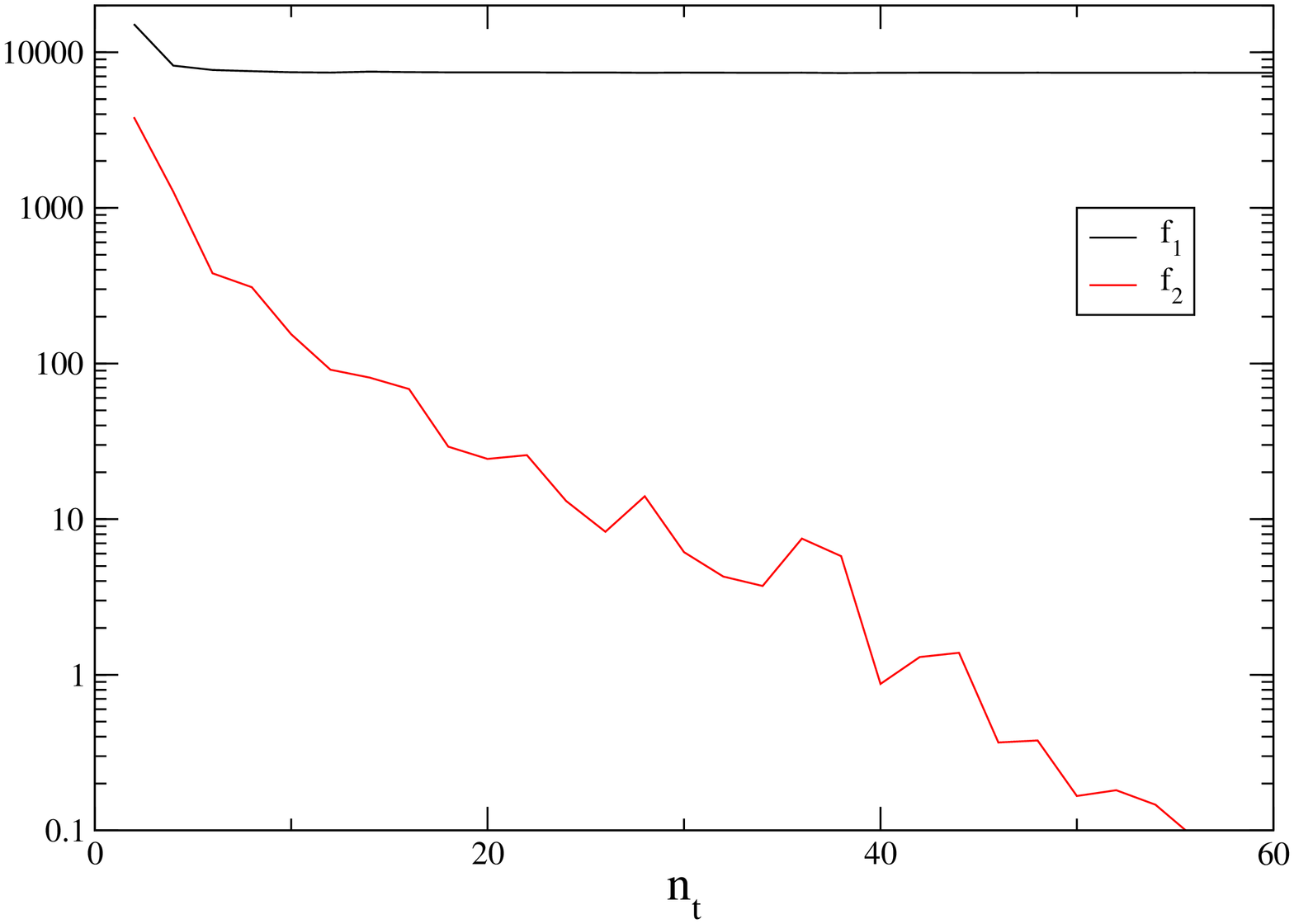}}}
\caption{$f_1$ and $f_2$ (see Eq.~(\protect\ref{eq:varia}))
as functions of $n_{\rm t}$ for $\Gamma=\gamma_3\gamma_5$ at $\kappa_{\rm loop}=0.166$, using the CG solver (left)
and the BiCGstab2 solver (right).\label{f1_f2_chan12}}
\end{figure}

\begin{figure}
\centerline{\rotatebox{270}{\includegraphics[height=.48\textwidth,clip]{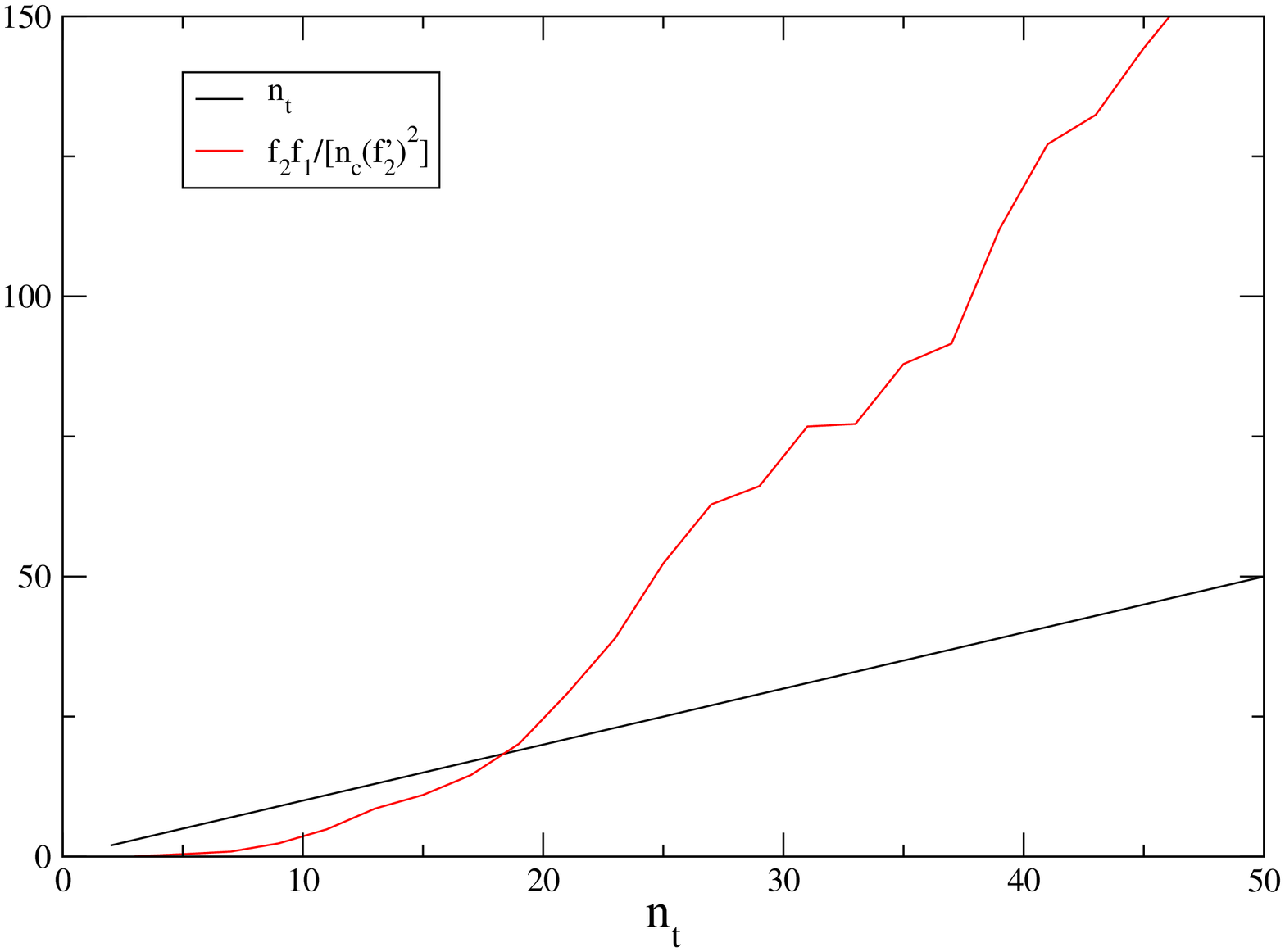}}\hspace{.02\textwidth}
\rotatebox{270}{\includegraphics[height=.495\textwidth,clip]{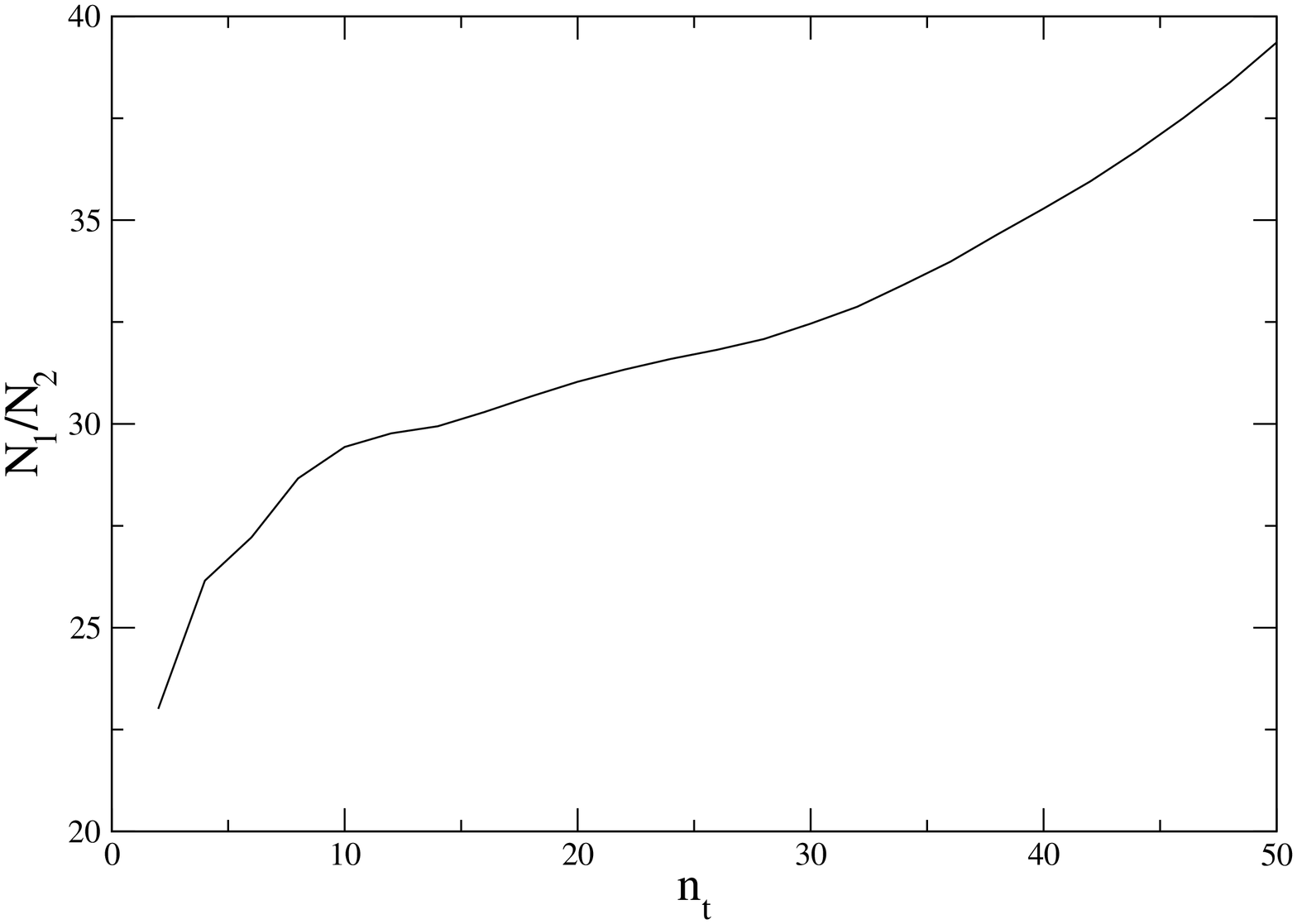}}}
\caption{Results used for the calculation of the optimal $n_{\rm t}$ and
$N_1/N_2$ values for $\Gamma=\gamma_3\gamma_5$ at $\kappa_{\rm loop}=0.166$.
In the left figure both sides of Eq.~(\protect\ref{opt_n}) are shown
and in the right figure
Eq.~(\protect\ref{opt_l1l2}).\label{fig_opt2}}
\end{figure}

\begin{table}
\begin{center}
\begin{tabular}{lccccccccccc}
\hline
$n_{\rm t}$ & 6 & 8 & 10 & 12 & 14 & 16 & 18 & 20 & 22 & 24 \\\hline
\multicolumn{11}{c}{$\Gamma = \mathbb{1}$}\\\hline
$N_1/N_2$ &7 &8 &2  &18  &57 & 35 &72 &110 &224 &303 \\
gain  &2.3  & 2.5 & 0.3 &2.0  &4.8  &4.8 &5.0 &3.4 &3.8 &3.0\\\hline
\multicolumn{11}{c}{$\Gamma = \gamma_3\gamma_5$}\\\hline
$N_1/N_2$ &20 &24  &48 &81 &93 &109  &254 &305 &288 &566 \\
gain  &6.1 &6.2 &4.4 &5.2 &7.2 &7.3 &6.9 &6.7 &6.8 &4.8 \\\hline
\end{tabular}
\end{center}
\caption{Gains obtained for various $n_{\rm t}$, setting
$N_1/N_2=f_1/f_2$, at 
  $\kappa_{\rm loop}=0.166$, using the BiCGstab2 solver.\label{gain_bicg}}
\end{table}

\appendix
\section{TSM parameter tuning for the CG and BiCGstab2 solvers}
\label{sec:app1}
The truncated solver method of Sec.~\ref{sec:tsm}
depends on two parameters, namely on
the number of iterations for the inexact solves, 
$n_{\rm t}$, and on the ratio
$N_1/N_2$ of the number of inexact estimates over the number of
estimates of the correction term, see Eq.~(\ref{trunc}).
These parameters need to be fixed, ideally, so as to minimize the
variance of the estimate of the disconnected loop,
$\mathrm{E}[\mathrm{Tr}\,(\mathrm{M}^{-1}\Gamma)]$, at fixed cost.
The estimates are uncorrelated and
for $N_1,N_2\gg 1$ the variance factorizes into,
\begin{equation}
\label{eq:varia}
\mathrm{var}[\mathrm{Tr}\,(\mathrm{M}^{-1}\Gamma)]=
\mathrm{var}\left(\overline{\langle\eta|\gamma_5\Gamma|s_{(n_{\rm t})}\rangle}^{N_1}\right)
+\mathrm{var}\left(\overline{\langle\eta|\gamma_5\Gamma(\,|s\rangle-|s_{(n_{\rm t})}\rangle)}^{N_2}\right)=
\frac{f_1}{N_1}+\frac{f_2}{N_2}\,,
\end{equation}
where $f_1$ and $f_2$ depend on $n_{\rm t}$ and $\Gamma$. An example of the
dependence on $n_{\rm t}$ is shown in Fig.~\ref{f1_f2_chan12} for
$\Gamma=\gamma_3\gamma_5$ for the CG and the BiCGstab2 solvers.
$f_1$ is roughly
independent of $n_{\rm t}$ and $f_2$ decreases rapidly with increasing
 $n_{\rm t}$. This behaviour is expected since after a few iterations
the first term of
Eq.~(\ref{trunc})  contains most of the signal
(and its error) while the second term (and its error) approaches zero.
The results were generated on a single configuration 
at $\kappa_{\rm loop}=0.166$ using
$300$ stochastic sources. Convergence is achieved after $n_{\rm conv}\approx
480$ CG iterations.

The approximate cost is given by,
\begin{equation}
\mathrm{cost} \approx N_1 n_{\rm t} + N_2 n_{\rm conv}\,.\label{costtsm}
\end{equation}
When TSM is combined with
HPE and/or TEA, there are corrections to this formula which the reader
can easily work out.
Using Lagrange multipliers, we minimize
the variance Eq.~(\ref{eq:varia}) at fixed cost,
assuming $f_1$ to be approximately
independent of $n_{\rm t}$.
This yields the optimal values,
\begin{align}
n_{\rm t}^{\rm opt} & =
\frac{1}{n_{\rm conv}} \frac{f_1 f_2}{{f_2'}^2}\,,\label{opt_n}\\
\frac{N_1}{N_2} & =  \sqrt{\frac{f_1}{f_2}
\frac{n_{\rm conv}}{n_{\rm t}^{\rm opt}} }\,,\label{opt_l1l2}
\end{align}
where $f_2'=d\!f_2 / dn_{\rm t}$. 

The right hand sides
of Eqs.~(\ref{opt_n}) and (\ref{opt_l1l2}) can be computed as
functions of $n_{\rm t}$ on one configuration, using a single set of
stochastic sources. By finding the intersection between the
curve $f_1f_2/(n_{\rm conv}{f_2^\prime}^2)$ and $n_{\rm t}$, one can extract
$n_{\rm t}^{\rm opt}$
and subsequently determine $N_1/N_2$ for this $n_{\rm t}$.
Fig.~\ref{fig_opt2} illustrates this procedure
for $\Gamma=\gamma_3\gamma_5$ for the CG solver. We read off the values
$n_{\rm t}^{\rm opt}\approx 18$ and $N_1/N_2\approx 30$ from this
figure, see Table~\ref{table:optvalues1}.

For our observables we find that when using the optimized TSM parameter values,
the two variances within Eq.~(\ref{eq:varia}) are of similar sizes,
$f_1/N_1\approx f_2/N_2$. It also turns out that the gain factor does not
critically depend on $N_1/N_2$. For instance, increasing this ratio
from the optimal value 
of 30 found for $\Gamma=\gamma_3\gamma_5$ at $n_{\rm t}=18$ to the equal
cost value $N_1/N_2=f_1/f_2\approx 35$ will increase the
final variance by just 3\,\%, which is statistically insignificant. 
This suggests an alternative
criterium for fixing the parameters: scanning
through $n_{\rm t}$, keeping
$N_1/N_2=f_1/f_2$ fixed, to
determine the $n_{\rm t}$ value with the smallest combined variance.
We find that following this strategy reduces no gain factor by more than
2\,\%, relative to the gain achieved using the optimal values,
at $\kappa=0.166$, using the CG solver.

It can be seen from Fig.~\ref{f1_f2_chan12} that the convergence
is no longer a smooth function of $n_{\rm t}$ when using the BiCGstab2
solver, so that $f_2^\prime$ cannot be determined. The situation
is even worse for $\Gamma=\mathbb{1}$.
In the BiCGstab2 case we have to resort to the alternative method discussed
above.
Table~\ref{gain_bicg} demonstrates that, using this approach,
one can indeed find values $n_{\rm t}$ and $N_1/N_2$ for the BiCGstab2 solver
that give reasonable gains. The cost was kept fixed to correspond to
300 BiCGstab2 solves to convergence.
In this case $n_{\rm conv}\approx 156$ is by a factor three
smaller than for the CG, however, each iteration is about
twice as expensive.
The best gain factors are 5.0 for $\Gamma=\mathbb{1}$ and
7.3 for $\Gamma=\gamma_3\gamma_5$ while for CG, using this
method, we are able
to achieve very similar factors of 4.9 and 7.9, respectively.

The CG algorithm is more robust than BiCGstab2 and gives 
nearly optimal gains over a wider range of $n_{\rm t}$ values.
Note for instance the somewhat erratic behaviour 
in Table~\ref{gain_bicg} of BiCGstab2 at
$n_{\rm t}=10$. Moreover,
the optimal
$N_1/N_2$ ratios come out rather large, due to tiny $f_2$ values,
which turns
BiCGstab2 less optimal when it is combined with HPE or TEA.
Therefore, in the context of TSM, the CG solver is our
preferred choice. However,
it is possible to combine both solvers, using CG for
the truncated solves and BiCGstab2 for running to convergence
more efficiently.


\begin{thebibliography}{00}
\bibitem{Jansen:2008wv}
  K.~Jansen, C.~Michael and C.~Urbach  [ETM Collaboration],
  The $\eta'$ meson from Lattice QCD,
  Eur.\ Phys.\ J.\  C {58} (2008) 261,
  \href{http://arxiv.org/abs/0804.3871}{arXiv:0804.3871 [hep-lat]}.
\bibitem{Bulava:2008qx}
  J.~Bulava, R.~Edwards, K.~J.~Juge, C.~J.~Morningstar and M.~J.~Peardon
                  [Hadron Spectrum Collaboration],
  Multi-hadron operators with all-to-all quark propagators,
PoS (LATTICE 2008) 124,
  \href{http://arxiv.org/abs/0810.0730}{arXiv:0810.0730 [hep-lat]}.
\bibitem{Aoki:2007rd}
  S.~Aoki, M.~Fukugita, K-I.~Ishikawa, N.~Ishizuka, K.~Kanaya, Y.~Kuramashi,
Y.~Namekawa, M.~Okawa, K.~Sasaki, A.~Ukawa, T. Yoshi\'e
  [CP-PACS Collaboration],
  Lattice QCD calculation of the $\rho$ meson decay width,
  Phys.\ Rev.\  D {76} (2007) 094506,
  \href{http://arxiv.org/abs/0708.3705}{arXiv:0708.3705 [hep-lat]}.
\bibitem{Zanotti:2008zm}
  J.~M.~Zanotti,
  Investigations of hadron structure on the lattice,
PoS (LATTICE 2008) 007,\\  
\href{http://arxiv.org/abs/0812.3845}{arXiv:0812.3845 [hep-lat]}.
\bibitem{Deka:2008xr}
  M.~Deka  T.~Streuer, T.~Doi, S.~J.~Dong, T.~Draper, K.-F.~Liu, N.~Mathur
 and A.~W.~Thomas,
  Moments of nucleon's parton distribution for the sea and valence quarks
  from Lattice QCD,
  Phys.\ Rev.\  D 79 (2009) 094502,
  \href{http://arxiv.org/abs/0811.1779}{arXiv:0811.1779 [hep-ph]}.
\bibitem{DeGrand:2006zz}
  T.~DeGrand and C.~DeTar,
  Lattice methods for quantum chromodynamics,
(World Scientific, Singapore, 2006).
\bibitem{Airapetian:2008qf}
  A.~Airapetian {\it et al.}  [HERMES Collaboration],
  Measurement of parton distributions of strange quarks in the nucleon from
  charged-kaon production in deep-inelastic scattering on the deuteron,
  Phys.\ Lett.\  B {666} (2008) 446,
  \href{http://arxiv.org/abs/0803.2993}{arXiv:0803.2993 [hep-ex]}.
\bibitem{Zhu:2002tn}
  S.~L.~Zhu, G.~Sacco and M.~J.~Ramsey-Musolf,
  Recoil order chiral corrections to baryon octet axial currents and large
  $N_c$ QCD,
  Phys.\ Rev.\  D {66} (2002) 034021,
\href{http://arXiv.org/abs/hep-ph/0201179}{arXiv:hep-ex/0201179}.
\bibitem{deFlorian:2009vb}
  D.~de Florian, R.~Sassot, M.~Stratmann and W.~Vogelsang,
  Extraction of spin-dependent parton densities and their uncertainties,
  Phys.\ Rev.\  D {80} (2009) 034030,
  \href{http://arxiv.org/abs/0904.3821}{arXiv:0904.3821 [hep-ph]}.
\bibitem{Airapetian:2007mh}
  A.~Airapetian {\it et al.}  [HERMES Collaboration],
  Precise determination of the spin structure function $g_1$ of the
 proton, deuteron and neutron,
  Phys.\ Rev.\ D {75} (2007) 012007,
\href{http://arXiv.org/abs/hep-ex/0609039}{arXiv:hep-ex/0609039}.
\bibitem{Frommer:1994vn}
  A.~Frommer, V.~Hannemann, B.~N\"ockel, T.~Lippert and K.~Schilling,
  Accelerating Wilson fermion matrix inversions by means of the stabilized
  biconjugate gradient algorithm,''
  Int.\ J.\ Mod.\ Phys.\  C {5} (1994) 1073,
\href{http://arXiv.org/abs/hep-lat/9404013}{arXiv:hep-lat/9404013}.
\bibitem{Collins:2007mh}
  S.~Collins, G.~S.~Bali and A.~Sch\"afer,
  Disconnected contributions to hadronic structure: a new method for
  stochastic noise reduction,
  PoS (LATTICE 2007) 141,
  \href{http://arXiv.org/abs/0709.3217}{arXiv:0709.3217 [hep-lat]}.
\bibitem{Bali:2008sx}
  G.~S.~Bali, S.~Collins and A.~Sch\"afer,
  Hunting for the strangeness content of the nucleon,
  PoS (LATTICE 2008) 161,
  \href{http://arXiv.org/abs/0811.0807}{arXiv:0811.0807 [hep-lat]}.
\bibitem{Wilson:1975vw}
  K.~G.~Wilson,
  Quarks on a lattice,
in Proceedings of
Methods in Field Theory, Les Houches 1975, Amsterdam 1976, 261;\\
Quarks on a lattice, or, the colored string model,
  Phys.\ Rept.\  {23} (1976) 331.
\bibitem{Bitar:1988bb}
  K.~Bitar, A.~D.~Kennedy, R.~Horsley, S.~Meyer and P.~Rossi,
  The QCD finite temperature transition and hybrid Monte Carlo,
  Nucl.\ Phys.\  B {313} (1989) 348.
\bibitem{z2min} M.~F.~Hutchinson,
A stochastic estimator of the trace of the influence matrix
for Laplacian smoothing splines,
Commun.\ Statist.-Simula.\ {18(3)} (1989) 1059.
\bibitem{Dong:1991xb}
  S.~J.~Dong and K.-F.~Liu,
  Quark loop calculations,
  Nucl.\ Phys.\ B (Proc.\ Suppl.)  {26} (1992) 353.
\bibitem{Bernardson:1993yg}
  S.~Bernardson, P.~McCarty and C.~Thron,
  Monte Carlo methods for estimating linear combinations of inverse matrix
entries in Lattice QCD,
  Comput.\ Phys.\ Commun.\ 78 (1993) 256.
\bibitem{Babich:2007jg}
  R.~Babich, R.~Brower, M.~Clark, G.~Fleming, J.~Osborn and C.~Rebbi,
  Strange quark contribution to nucleon form factors,
  PoS (LATTICE 2007) 139,
  \href{http://arXiv.org/abs/0710.5536}{arXiv:0710.5536 [hep-lat]}.
\bibitem{Wilcox:1999ab}
  W.~Wilcox,
  Noise methods for flavor singlet quantities,
  \href{http://arXiv.org/abs/hep-lat/9911013}{arXiv:hep-lat/9911013}.
\bibitem{Viehoff:1997wi}
  J.~Viehoff, N.~Eicker, S.~G\"usken, H.~Hoeber, P.~Lacock, T.~Lippert, G.~Ritzenh\"ofer, K.~Schilling, A.~Spitz, P. Ueberholz [SESAM Collaboration],
  Improving stochastic estimator techniques for disconnected diagrams,
  Nucl.\ Phys.\ B (Proc.\ Suppl.)  {63} (1998) 269,
  \href{http://arXiv.org/abs/hep-lat/9710050}{arXiv:hep-lat/9710050}.
\bibitem{Foley:2005ac}
A.~O'Cais, J.~Foley, K.~Jimmy Juge, M.~Peardon, S.~M.~Ryan and J.~I.~Skullerud
[TrinLat Collaboration],
Improving algorithms to compute all elements of the lattice quark
propagator,
Nucl.\ Phys.\ B (Proc.\ Suppl.)  {140} (2005) 844,
  \href{http://arXiv.org/abs/hep-lat/0409069}{arXiv:hep-lat/0409069};\\
  Practical all-to-all propagators for Lattice QCD,
  Comput.\ Phys.\ Commun.\  {172} (2005) 145,
  \href{http://arXiv.org/abs/hep-lat/0505023}{arXiv:hep-lat/0505023}.
\bibitem{Thron:1997iy}
  C.~Thron, S.~J.~Dong, K.~F.~Liu and H.~P.~Ying,
  Pad\'e-$Z_2$ estimator of determinants,
  Phys.\ Rev.\ D 57 (1998) 1642,
  \href{http://arXiv.org/abs/hep-lat/9707001}{arXiv:hep-lat/9707001}.
\bibitem{Michael:1999rs}
  C.~Michael, M.~S.~Foster and C.~McNeile  [UKQCD Collaboration],
  Flavour-singlet pseudoscalar and scalar mesons,
  Nucl.\ Phys.\ B (Proc.\ Suppl.)  {83} (2000) 185,
  \href{http://arXiv.org/abs/hep-lat/9909036}{arXiv:hep-lat/9909036}.
\bibitem{Bali:2005pb}
  G.~S.~Bali, T.~D\"ussel, T.~Lippert, H.~Neff, Z.~Prkacin and K.~Schilling
[SESAM Collaboration],
  String breaking with dynamical Wilson fermions,
  Nucl.\ Phys.\ B (Proc.\ Suppl.)  {140} (2005) 609,
  \href{http://arXiv.org/abs/hep-lat/}{arXiv:hep-lat/0409137};\\
  G.~S.~Bali, T.~D\"ussel, T.~Lippert, H.~Neff and K.~Schilling
[SESAM Collaboration],
  Observation of string breaking in QCD,
  Phys.\ Rev.\  D {71} (2005) 114513,
  \href{http://arXiv.org/abs/hep-lat/0505012}{arXiv:hep-lat/0505012}.
\bibitem{DeGrand:2002gm}
  T.~A.~DeGrand and U.~M.~Heller  [MILC Collaboration],
  Witten-Veneziano relation, quenched QCD, and overlap fermions,
  Phys.\ Rev.\  D {65} (2002) 114501,
  \href{http://arXiv.org/abs/hep-lat/0202001}{arXiv:hep-lat/0202001}.
\bibitem{Morgan:2004zh}
  R.~B.~Morgan and W.~Wilcox,
  Deflated iterative methods for linear equations with multiple right-hand
  sides,
  \href{http://arXiv.org/abs/math-ph/0405053}{arXiv:math-ph/0405053}.
\bibitem{Luscher:2007se}
  M.~L\"uscher,
  Local coherence and deflation of the low quark modes in Lattice QCD,
  JHEP {0707} (2007) 081,
  \href{http://arXiv.org/abs/0706.2298}{arXiv:0706.2298 [hep-lat]}.
\bibitem{Stathopoulos:2007zi}
  A.~Stathopoulos and K.~Orginos,
  Computing and deflating eigenvalues while solving multiple right hand side
  linear systems in Quantum Chromodynamics,
  \href{http://arXiv.org/abs/0707.0131}{arXiv:0707.0131 [hep-lat]}.
\bibitem{Brannick:2007ue}
  J.~Brannick, R.~C.~Brower, M.~A.~Clark, J.~C.~Osborn and C.~Rebbi,
  Adaptive multigrid algorithm for Lattice QCD,
  Phys.\ Rev.\ Lett.\  {100} (2008) 041601,
  \href{http://arXiv.org/abs/0707.4018}{arXiv:0707.4018 [hep-lat]}.
\bibitem{Wilcox:2007ei}
  W.~M.~Wilcox,
  Deflation methods in fermion inverters,
  PoS (LATTICE 2007) 025,
  \href{http://arXiv.org/abs/0710.1813}{arXiv:0710.1813 [hep-lat]}.
\bibitem{Tadano:2009gg}
  H.~Tadano, Y.~Kuramashi and T.~Sakurai,
  Application of preconditioned block BiCGGR to the Wilson-Dirac equation
  with multiple right-hand sides in Lattice QCD,
  \href{http://arXiv.org/abs/0907.3261}{arXiv:0907.3261 [hep-lat]}.
\bibitem{Ehmann:2009ki}
  C.~Ehmann and G.~S.~Bali,
  $\eta'$-$\eta_c$-mixing with improved stochastic estimators,
  PoS (LATTICE 2008) 114,
  \href{http://arXiv.org/abs/0903.2947}{arXiv:0903.2947 [hep-lat]}.
\bibitem{Gusken:1999te}
  S.~G\"usken,
  Flavor singlet phenomena in Lattice QCD,
  \href{http://arXiv.org/abs/hep-lat/9906034}{arXiv:hep-lat/9906034}.
\bibitem{Neuberger:1997fp}
  H.~Neuberger,
  Exactly massless quarks on the lattice,
  Phys.\ Lett.\  B {417} (1998) 141,
  \href{http://arXiv.org/abs/hep-lat/9707022}{arXiv:hep-lat/9707022}.
\bibitem{Sheikholeslami:1985ij}
  B.~Sheikholeslami and R.~Wohlert,
  Improved continuum limit lattice action for QCD with Wilson fermions,
  Nucl.\ Phys.\  B {259} (1985) 572.
\bibitem{Mathur:2002sf}
  N.~Mathur and S.~J.~Dong,
 Study of stochastic estimates of quark loops with unbiased subtraction,
 Nucl.\ Phys.\ B (Proc.\ Suppl.) 119 (2003) 401,
  \href{http://arXiv.org/abs/hep-lat/0209055}{arXiv:hep-lat/0209055}.
\bibitem{Eicker:1996gk}
  N.~Eicker, U.~Gl\"assner, S.~G\"usken, H.~Hoeber, T.~Lippert, G.~Ritzenh\"ofer, K.~Schilling, G.~Siegert. A.~Spitz, P.~Ueberholz and J.~Viehoff
[SESAM Collaboration],
  Evaluating sea quark contributions to flavour-singlet operators in  lattice
  QCD,
  Phys.\ Lett.\  B {389} (1996) 720,
  \href{http://arXiv.org/abs/hep-lat/9608040}{arXiv:hep-lat/9608040}.
\bibitem{Neff:2001zr}
  H.~Neff, N.~Eicker, T.~Lippert, J.~W.~Negele and K.~Schilling,
  On the low fermionic eigenmode dominance in QCD on the lattice,
  Phys.\ Ref.\ D 64 (2001) 114509,
  \href{http://arXiv.org/abs/hep-lat/0106016}{arXiv:hep-lat/0106016}.
\bibitem{Kalkreuter:1995mm}
  T.~Kalkreuter and H.~Simma,
  An accelerated conjugate gradient algorithm to compute low lying
 eigenvalues: a study for the Dirac operator in $\mathrm{SU}(2)$ Lattice QCD,
  Comput.\ Phys.\ Commun.\  {93} (1996) 33
  \href{http://arXiv.org/abs/hep-lat/9507023}{arXiv:hep-lat/9507023}.
\bibitem{DeGrand:2004qw}
  T.~A.~DeGrand and S.~Sch\"afer,
  Improving meson two-point functions in Lattice QCD,
  Comput.\ Phys.\ Commun.\ 159 (2004) 185,
  \href{http://arXiv.org/abs/hep-lat/0401011}{arXiv:hep-lat/0401011}.
\bibitem{Giusti:2004yp}
  L.~Giusti, P.~Hern\'andez, M.~Laine, P.~Weisz and H.~Wittig,
 Low-energy couplings of QCD from current correlators near the chiral
  limit, JHEP 04 (2004) 013,
  \href{http://arXiv.org/abs/hep-lat/0402002}{arXiv:hep-lat/0402002}.
\bibitem{deForcrand:1995bs}
  P.~de Forcrand,
  Progress on lattice QCD algorithms,
  Nucl.\ Phys.\ B (Proc.\ Suppl.)\  {47} (1996) 228,\\
  \href{http://arXiv.org/abs/hep-lat/9509082}{arXiv:hep-lat/9509082}.
\bibitem{Darnell:2007dr}
  D.~Darnell, R.~B.~Morgan and W.~Wilcox,
  Deflated GMRES for systems with multiple shifts and multiple right-hand
  sides,
  Linear Algebra Appl.\  429 (2008) 2415,
  \href{http://arXiv.org/abs/0707.0502}{arXiv:0707.0502 [math-ph]}.
\bibitem{Hip:2001hc}
  I.~Hip, T.~Lippert, H.~Neff, K.~Schilling and W.~Schroers,
  Instanton dominance of topological charge fluctuations in QCD?,
  Phys.\ Rev.\  D {65} (2002) 014506,
  \href{http://arXiv.org/abs/hep-lat/0105001}{arXiv:hep-lat/0105001}.
\bibitem{Michael:1998sg}
  C.~Michael and J.~Peisa  [UKQCD Collaboration],
  Maximal variance reduction for stochastic propagators with applications  to
  the static quark spectrum,
  Phys.\ Rev.\  D {58} (1998) 034506,
  \href{http://arXiv.org/abs/hep-lat/9802015}{arXiv:hep-lat/9802015}.
\bibitem{Burch:2006mb}
  T.~Burch and C.~Hagen,
  Domain decomposition improvement of quark propagator estimation,
  Comput.\ Phys.\ Commun.\  {176} (2007) 137,
  \href{http://arXiv.org/abs/hep-lat/0607029}{arXiv:hep-lat/0607029}.
\bibitem{Foster:1998vw}
  M.~Foster and C.~Michael  [UKQCD Collaboration],
  Quark mass dependence of hadron masses from Lattice QCD,
  Phys.\ Rev.\  D {59} (1999) 074503,
  \href{http://arXiv.org/abs/hep-lat/9810021}{arXiv:hep-lat/9810021}.
\bibitem{McNeile:2006bz}
  C.~McNeile and C.~Michael  [UKQCD Collaboration],
  Decay width of light quark hybrid meson from the lattice,
  Phys.\ Rev.\  D 73 (2006) 074506,
  \href{http://arXiv.org/abs/hep-lat/0603007}{arXiv:hep-lat/0603007}.
\bibitem{Boyle:2008yd}
  P.~A.~Boyle,  J.~M.~Flynn, A.~J\"uttner, C.~Kelly, H.~Pedroso de Lima,
C.~M.~Maynard, C.~T.~Sachrajda and J.~M.~Zanotti [UKQCD Collaboration],
  The pion's electromagnetic form factor at small momentum transfer in full
  Lattice QCD, JHEP 0807 (2008) 112,
  \href{http://arXiv.org/abs/0804.3971}{arXiv:0804.3971 [hep-lat]}.
\bibitem{Duncan:2001ta}
  A.~Duncan and E.~Eichten,
  Improved pseudofermion approach for all-point propagators,
  Phys.\ Rev.\  D {65} (2002) 114502,
  \href{http://arXiv.org/abs/hep-lat/0112028}{arXiv:hep-lat/0112028}.
\bibitem{deForcrand:1998je}
  P.~de Forcrand,
  Monte Carlo quasi-heatbath by approximate inversion,
  Phys.\ Rev.\  E 59 (1999) 3698,\\
  \href{http://arXiv.org/abs/cond-mat/9811025}{arXiv:cond-mat/9811025}.
  \bibitem{latdetails} Y.~Aoki, Z.~Fodor, S.~D.~Katz and K.~K.~Szabo,
    The equation of state in Lattice QCD: with physical quark
      masses towards the continuum limit,
JHEP 01 (2006) 089,
    \href{http://arXiv.org/abs/hep-lat/0510084}{arXiv:hep-lat/0510084}.
\bibitem{chroma1} R.~G.~Edwards [LHP Collaboration] and B.~Jo\'o [UKQCD
    Collaboration], The Chroma software system for Lattice
      QCD,
  Nucl.\ Phys.\ B (Proc.\ Suppl.)  {140} (2005) 832,
  \href{http://arXiv.org/abs/hep-lat/0409003}{arXiv:hep-lat/0409003}.
\bibitem{chroma2}
C.~McClendon, Optimized Lattice
      QCD kernels for a Pentium 4 cluster,
Jlab preprint (2001)
    JLAB-THY-01-29,
\href{http://www.jlab.org/~edwards/qcdapi/reports/dslash_p4.pdf}{http://www.jlab.org/$\sim$edwards/qcdapi/reports/dslash\_p4.pdf}.
\bibitem{chroma3} P.A.~Boyle, \href{http://www.ph.ed.ac.uk/~paboyle/bagel/Bagel.html}{http://www.ph.ed.ac.uk/$\sim$paboyle/bagel/Bagel.html} (2005).
\bibitem{DelDebbio:2005qa}
  L.~Del Debbio, L.~Giusti, M.~L\"uscher, R.~Petronzio and N.~Tantalo,
  Stability of Lattice QCD simulations and the thermodynamic limit,
  JHEP 0602 (2006) 011,
  \href{http://arXiv.org/abs/hep-lat/0512021}{arXiv:hep-lat/0512021}.
\bibitem{Skouroupathis:2008mf}
  A.~Skouroupathis and H.~Panagopoulos,
  Two-loop renormalization of vector, axial-vector and tensor fermion
  bilinears on the lattice,
  Phys.\ Rev.\  D {79} (2009) 094508,
  \href{http://arXiv.org/abs/0811.4264}{arXiv:0811.4264 [hep-lat]}.
\bibitem{Babich:2009rq}
  R.~Babich, R.~Brower, M.~Clark, G.~Fleming, J.~Osborn and C.~Rebbi,
  Strange quark content of the nucleon,
  PoS (LATTICE 2008) 160,
  \href{http://arXiv.org/abs/0901.4569}{arXiv:0901.4569 [hep-lat]}.
\bibitem{Fukugita:1994fh}
  M.~Fukugita, Y.~Kuramashi, M.~Okawa and A.~Ukawa,
  Proton spin structure from Lattice QCD,
  Phys.\ Rev.\ Lett.\  {75} (1995) 2092,
  \href{http://arXiv.org/abs/hep-lat/9501010}{arXiv:hep-lat/9501010}.
\bibitem{Dong:1995rx}
  S.~J.~Dong, J.~F.~Lagae and K.-F.~Liu,
  Flavor singlet $g_A$ from Lattice QCD,
  Phys.\ Rev.\ Lett.\  {75} (1995) 2096,
  \href{http://arXiv.org/abs/hep-ph/9502334}{arXiv:hep-ph/9502334}.
\bibitem{Gusken:1999as}
  S.~G\"usken, J.~Viehoff, N.~Eicker, T.~Lippert, K.~Schilling, A.~Spitz
and T.~Struckmann [T$\chi$L Collaboration],
  Flavor singlet axial vector coupling of the proton with dynamical Wilson
  fermions,
  Phys.\ Rev.\  D {59} (1999) 114502,
  \href{http://arXiv.org/abs/hep-lat/9901009}{arXiv:hep-lat/9901009}.
\bibitem{inprep}
  G.~S.~Bali, S.~Collins and A.~Sch\"afer  [QCDSF Collaboration],
  Strangeness and charm content of the nucleon,
PoS (LAT 2009) 149,
\href{http://arXiv.org/abs/0911.2407}{arXiv:0911.2407 [hep-lat]}.
\end{thebibliography}
\end{document}